\documentclass[pre,aps,twocolumn,showpacs,floatfix]{revtex4-1}
\usepackage{graphicx}
\usepackage{color}
\usepackage{amsmath, amsthm, amssymb}
\newcommand{\be}{\begin{equation}}
\newcommand{\ee}{\end{equation}}
\newcommand{\bea}{\begin{eqnarray}}
\newcommand{\eea}{\end{eqnarray}}

\begin{document}

\title{Phase Transitions in a System of Hard Rectangles on the Square Lattice}
\author{Joyjit Kundu}
\email{joyjit@imsc.res.in}
\affiliation{The Institute of Mathematical Sciences, C.I.T. Campus,
Taramani, Chennai 600113, India}
\author{R. Rajesh}
\email{rrajesh@imsc.res.in}
\affiliation{The Institute of Mathematical Sciences, C.I.T. Campus,
Taramani, Chennai 600113, India}

\date{\today}

\begin{abstract}

The phase diagram of a system of monodispersed hard rectangles of size 
$m\times m k$ on a square lattice is numerically determined for $m=2,3$ 
and aspect ratio $k= 1,2,\ldots, 7$. We show the existence of a 
disordered phase, a nematic phase with orientational order, a columnar 
phase with orientational and partial translational order, and a 
solid-like phase with sublattice order, but no orientational order. 
The asymptotic behavior of the phase 
boundaries for large $k$ are determined using a combination of entropic 
arguments and a Bethe approximation. This allows us to generalize the 
phase diagram to larger $m$ and $k$, showing that for $k \geq 7 $, the 
system undergoes three entropy driven phase transitions with increasing 
density. The nature of the different phase transitions are established 
and the critical exponents for the continuous transitions are determined 
using finite size scaling.

\end{abstract}


\maketitle

\section{\label{sec:intro}Introduction}

The study of entropy-driven phase transitions in a system of long hard 
rods has a long history dating back to Onsager's 
demonstration~\cite{onsager1949} that the three dimensional system 
undergoes a transition from an isotropic phase to an orientationally 
ordered nematic phase as the density of the rods is 
increased~\cite{onsager1949,flory1956b,degennesBook,vroege1992,dhar2011}.  
Further increase in density may result in a smectic phase that partially 
breaks translational symmetry, and a solid 
phase~\cite{frenkel1988,frenkel1997}. In two dimensional continuum 
space, the continuous rotational symmetry remains unbroken but the 
system undergoes a Kosterlitz-Thouless type transition from a 
low-density phase with exponential decay of orientational correlations 
to a high-density phase having quasi long range 
order~\cite{straley1971,frenkel1985,frenkel2004,khandkar2005,donev2006,zhao2007,vink2009}. 
Experimental realizations include tobacco mosaic 
virus~\cite{fraden1989}, liquid crystals~\cite{degennesBook}, carbon 
nanotube gels~\cite{islam2004}, and brownian squares~\cite{zhao2011}. 
The phenomenology is, however, much less clear when the orientations are 
discrete, and the positions are either on a lattice or in the continuum, 
when even the existence of the nematic phase has been 
convincingly seen in simulations only 
recently~\cite{ghosh2007,giuliani2013}.

Consider hard rectangles of size $m \times m k$ on a two-dimensional 
square lattice where each rectangle occupies $m$ ($m k$) lattice sites 
along the short (long) axis. The limiting cases when either the aspect 
ratio $k=1$ or $m=1$ are better studied. When $m=1$ and $k\geq 7$, there 
are, remarkably, two entropy-driven transitions: from a low-density 
isotropic phase to an intermediate density nematic phase, and from the 
nematic phase to a high-density disordered 
phase~\cite{ghosh2007,joyjit2013}. While the first transition is in the 
Ising universality class~\cite{fernandez2008a,fischer2009}, the second 
transition could be non-Ising~\cite{joyjit2013}, and it is 
not yet understood 
whether the high density phase is a re-entrant low density phase or a 
new phase~\cite{joyjit2013,joyjit_rltl2013}. When $k=1$ (hard squares), 
the system undergoes a transition into a high density columnar phase. 
The transition is continuous for 
$m=2$~\cite{baxterBook,fernandes2007,hirokazu2007,nienhuis2011,kabirthesis}, 
and first order for $m=3$~\cite{fernandes2007}. When $m\rightarrow 
\infty$, keeping $k$ fixed, the lattice model is equivalent to the model 
of oriented rectangles in two dimensional continuum, also known as the 
Zwanzig model~\cite{zwanzig1963}. For oriented lines in the continuum 
($k\rightarrow \infty$), a nematic phase exists at high 
density~\cite{fischer2009}. The only theoretical results that exist are 
when $m=1$ and $k=2$ (dimers), for which no nematic phase 
exists~\cite{Heilmann1970,Kunz1970,Gruber1971,lieb1972}, $k \gg 1$, when 
the existence of the nematic phase may be proved 
rigorously~\cite{giuliani2013}, and an exact solution for arbitrary $k$ 
on a tree like lattice~\cite{dhar2011,joyjit_rltl2013}.

Less is known for other values of $m$ and $k$. Simulations of rectangles 
of size $2\times 5$ did not detect any phase transition with increasing 
density~\cite{barnes2009}, while those of parallelepipeds on cubic 
lattice show layered and columnar phases, but no nematic 
phase~\cite{casey1995}. In general, numerical studies of large 
rectangles are constrained by the fact that it is difficult to 
equilibrate the system at high densities using Monte Carlo algorithms 
with local moves, as the system gets jammed and requires 
correlated moves of several particles to access different configurations. 

In addition to being the lattice version of the hard rods problem, the 
study of lattice models of hard rectangles is useful in understanding 
the phase transitions in adsorbed monolayers on crystal surfaces. The 
$(100)$ and $(110)$ planes of fcc crystals have square and rectangular 
symmetry and may be treated with lattice statistics if the 
adsorbate-adsorbate interaction is negligible with respect to the 
periodic variation of the corrugation potential of the underlying 
substrate~\cite{binder2000}. For example, the critical behavior of a 
monolayer of chlorine (Cl) on Ag$(100)$ is well reproduced by the hard 
square model ($k=1$) and the high density $c(2\times 2)$ structure of 
the Cl-adlayer may be mapped to the high density phase of the hard 
square problem~\cite{taylor1985}. Structures such as $p(2\times 2)$, 
$c(2\times2)$ and $(2\times1)$ are ordered structures. Lattice gas 
model with repulsive interaction up to forth nearest neighbor has been 
used to study the phase behavior of selenium adsorbed on 
Ni$(100)$~\cite{bak1985}. The results in this paper show that different 
phases with orientational and positional order may be obtained by only 
hard core exclusion.

At a more qualitative level, discrete models of hard rods have been used 
to obtain realistic phase diagram for polydispersed 
systems~\cite{sear2000,cuesta2003}, to describe orientational wetting of 
rods~\cite{roij2000}, to model and understand self-assembly of nano 
particles on monolayers~\cite{barnes2009} and thermodynamics of linear 
adsorbates~\cite{pastor1999,pastor2003}.

Lattice models of hard rectangles also falls in the general class of 
hard core lattice gas models of differently shaped particles. The study 
of these models also have been of continued interest in Statistical 
Physics as minimal models for the melting transitions. The different 
shapes studied include squares~\cite{nigam1967,pearce1988,kabir2012}, 
hexagons on triangular~\cite{baxter1980,heilmann1973} and square 
lattices~\cite{dickman2012}, triangles~\cite{nienhuis1999}, and 
tetrominoes~\cite{barnes2009}.

In this paper, we adapt and implement an efficient Monte Carlo algorithm 
with cluster moves that was very effective in studying the hard rod 
($m=1$) problem on lattices~\cite{joyjit2013,joyjit_dae}. The hard 
rectangle model and the algorithm are described in detail in 
Sec.~\ref{sec:model}. We observe four distinct phases at different 
densities: isotropic, nematic, columnar, and sublattice phases. These 
phases, suitable order parameters to characterize them, and other 
thermodynamic quantities are defined in Sec.~\ref{sec:phases}. From 
extensive large scale simulations, we determine the rich phase diagram 
for $m=2, 3$, and $k=1,\ldots, 7$. The phase diagram for $m=2$ is 
discussed in Sec.~\ref{sec:phase_diag2}. We find that all transitions 
except the isotropic-columnar transition for $k=6$ are continuous. The 
critical exponents and universality classes of the continuous 
transitions are determined. 
Section~\ref{sec:phase_diag3} contains the details about the phase 
diagram and the nature of the phase transitions for $m=3$. In 
Sec.~\ref{sec:phase_boundary}, we use a Bethe approximation and 
estimates of  entropies for the different phases to determine the 
phase boundaries for large $k$, allowing us to generalize the phase 
diagram to arbitrary $m$ and $k$. In particular, it allows us to take 
the continuum limit $m \to \infty$, thus obtaining predictions for a
system of oriented 
rectangles in the continuum. Sec.~\ref{sec:summary} contains a summary 
and a discussion of possible extensions of the problem.

\section{\label{sec:model}The Model and Monte Carlo algorithm}

We define the model on a square lattice of size $L \times L$ with 
periodic boundary conditions. Consider a mono-dispersed system of 
rectangles of size $m \times m k$ such that the aspect ratio is $k$. A 
rectangle can be either horizontal or vertical. A horizontal (vertical) 
rectangle occupies $m k$ sites along the $x$ ($y$)-axis and $m$ sites 
along the $y$ ($x$)-axis. No two rectangles may overlap. An activity 
$z=e^\mu$ is associated with each rectangle, where $\mu$ is the chemical 
potential.
\begin{figure} 
\includegraphics[width=\columnwidth]{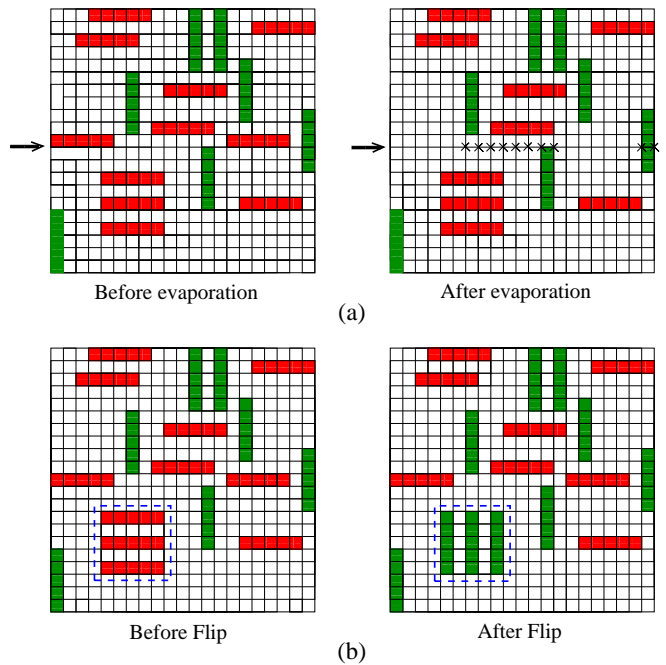} 
\caption{(Color online) An illustration of the Monte Carlo algorithm. 
(a) Configurations before and after the evaporation of horizontal 
rectangles with head on a particular row (denoted by an arrow). Sites 
denoted by cross symbols cannot be occupied by horizontal rectangles in 
the new configuration. (b) An example of the flip move for rectangles of 
size $2 \times 6$. A rotatable or flippable plaquette of size $6\times 
6$, consisting of three aligned rectangles, is shown by the dashed line. 
After the flip move, the horizontal rectangles become vertical.} 
\label{fig:flip} 
\end{figure}

We study this system using constant $\mu$ grand canonical Monte Carlo 
simulations. The algorithm that we implement is an adaptation of the 
algorithm with cluster moves that was introduced in 
Refs.~\cite{joyjit2013,joyjit_dae} to study the problem of hard rods 
(m=1). We describe the algorithm below. Given a valid configuration of 
rectangles, in a single move, a row or a column is chosen at random. If 
a row is chosen, then all horizontal rectangles whose heads (bottom, 
left corner) lie in that row are removed, leaving the rectangles with 
heads in other rows untouched. The emptied row now consists of two kinds 
of sites: forbidden sites that cannot be occupied with horizontal 
rectangles due to the presence of vertical rectangles in the same row or 
due to rectangles with heads in the neighboring $(m-1)$ rows, and sites 
that may be occupied by horizontal rectangles in a valid configuration. 
An example illustrating the forbidden sites is shown in 
Fig.~\ref{fig:flip}(a). It is clear that the sites that may be occupied 
are divided into intervals of contiguous empty sites. The problem of 
occupation of the emptied row with a new configuration now reduces to 
the problem of occupying the empty intervals. However, the empty 
intervals may be occupied
independent of each other, as the occupation of 
one is not affected by the configuration of rectangles in the remaining 
ones. Thus, the re-occupation of the emptied row reduces to a problem of 
occupying a one dimensional interval with rods. This problem is easily 
solvable and the equilibrium probabilities of each new configuration may 
be easily calculated. We refer to Refs.~\cite{joyjit2013,joyjit_dae} for 
the calculation of these probabilities.  If a column is chosen instead 
of a row, then a similar operation is performed for the vertical 
rectangles whose heads lie in that column. 

In addition to the above evaporation-deposition move, we find that the 
autocorrelation time is reduced considerably by introducing a flip move. 
In this move, a site $(i,j)$ is picked at random. If it is occupied by 
the head of a horizontal rectangle, then we check whether $(i,j+m), 
(i,j+2 m),\ldots, (i,j+[k-1] m)$ sites are occupied by the heads of 
horizontal rectangles. If that is the case, we call this set of $k$ aligned 
rectangles a rotatable plaquette of horizontal
rectangles. 
In the 
flip move such a rotatable plaquette 
of size $mk \times mk$, containing $k$ horizontal rectangles, is 
replaced by a similar plaquette of $k$ vertical rectangles. 
An example of the flip move is shown in 
Fig.~\ref{fig:flip}(b). If $(i,j)$ is occupied by the head of a vertical 
rectangle and a rotatable plaquette of vertical rectangles is present, 
then 
it is replaced by a plaquette of $k$ aligned horizontal rectangles. 
A Monte Carlo move 
corresponds to $2 L$ evaporation--deposition moves and $L^2$ flip moves. 
It is easy to check that the algorithm is ergodic and obeys detailed 
balance.

We implement a parallelized version of the above algorithm. In the 
evaporation-deposition move, we simultaneously update all rows that are 
separated by $m$. Once all rows are updated in this manner, the columns 
are updated. We also parallelize the flip move. The lattice is divided 
into $L^2/(m^2 k^2)$ blocks of size $m k\times m k$. The flipping of 
each of these blocks is independent of the other and may therefore be 
flipped simultaneously. We flip a rotatable plaquette with probability 
$1/2$. The parallelization and efficiency of the algorithm allows us to 
simulate large systems (up to $L=810$) at high densities (up to $0.99$).

We check for equilibration by starting the simulations with two 
different initial configurations and making sure that the final 
equilibrium state is independent of the initial condition. One 
configuration is a fully nematic state, where all rectangles are either 
horizontal or vertical and the other is a random configuration where 
rectangles of both vertical and horizontal orientations are deposited at 
random.

\section{\label{sec:phases}The Different Phases}

\begin{figure}
\includegraphics[width=\columnwidth]{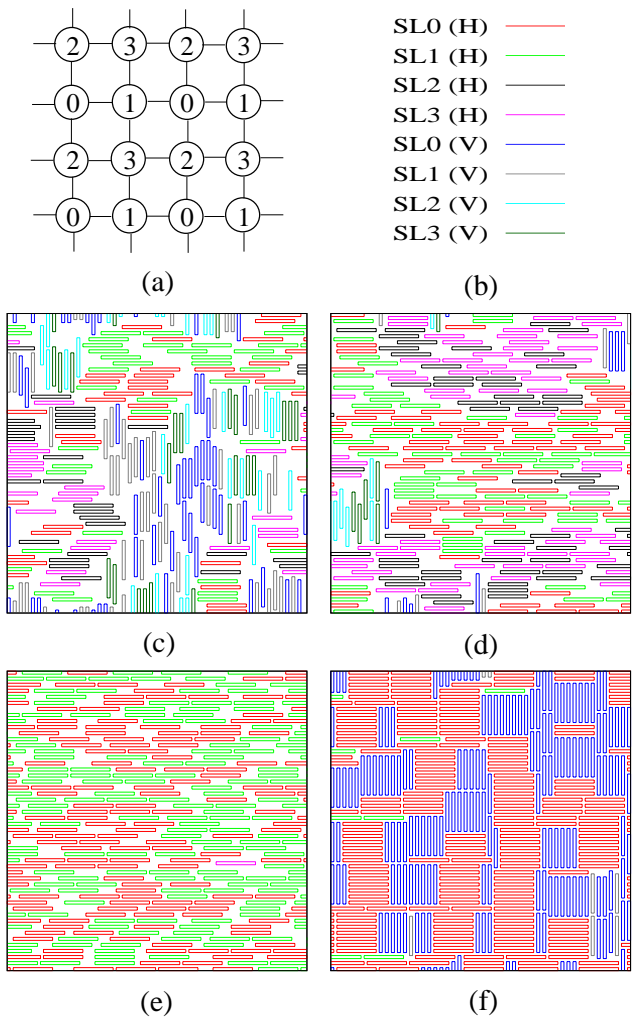}
\caption{(Color online) Snapshots of different phases (at different 
densities) in a system of $2\times 14$ hard rectangles. (a) The four sub 
lattices when $m=2$. (b) The color scheme: eight colors corresponding
to two orientations, horizontal (H) and vertical (V), and heads of 
rectangles being on one of the four sublattices, denoted by SL0 to SL3. 
(c) The isotropic phase where all $8$ colors are 
present. (d) The nematic phase, dominated by $4$ colors corresponding to 
4 sublattices and one orientation. (e) The columnar phase, dominated by 
$2$ colors corresponding to two sublattices and one orientation. (f) The 
sublattice phase, dominated by $2$ colors corresponding to one 
sublattice and $2$ orientations.} 
\label{fig:fig_snap} 
\end{figure}

Snapshots of the different phases that we observe in simulations are 
shown in Fig.~\ref{fig:fig_snap}. First is the low density isotropic (I) 
phase in which the rectangles have neither orientational nor 
translational order [see Fig.~\ref{fig:fig_snap}(c)]. Second is the 
nematic (N) phase in which the rectangles have orientational order but 
no translational order [see Fig.~\ref{fig:fig_snap}(d)] . In this phase, 
the mean number of horizontal rectangles is different from that of 
vertical rectangles. The third phase is the columnar (C) phase, having 
orientational order and partial translational order [see 
Fig.~\ref{fig:fig_snap}(e)]. In this phase, if majority of rectangles 
are horizontal (vertical), then their heads, or bottom left corners, 
preferably lie in rows (columns) that are separated by $m$. Thus, it 
breaks the translational symmetry in the direction perpendicular to the 
orientation but not parallel to the orientation.  Clearly, there are $2 
m$ symmetric C phases. In this phase the rectangles can slide 
much more along one lattice direction.
The fourth phase is the crystalline sublattice 
(S) phase with no orientational order [see Fig.~\ref{fig:fig_snap}(f)]. 
We divide the square lattice into $m^2$ sublattices by assigning to a 
site $(i,j)$ a label $(i \mod m) +m \times (j \mod m)$. The sublattice 
labeling for the case $m=2$ is shown in Fig.~\ref{fig:fig_snap}(a). In 
the S phase, the heads of the rectangles preferably occupy one 
sublattice, breaking translational symmetry in both the directions.

From the symmetry of the system we would expect up to $7$ phases. The 
orientational symmetry could be present or broken while the 
translational symmetry could be unbroken, broken along only one 
or both $x$- and $y$- directions.  
If the orientational symmetry is broken, then the translational 
symmetry could be broken either parallel or perpendicular to the 
preferred orientations. Out of the $7$ possibilities, we do not observe 
(i) a phase with no orientational order but partial translational order, 
(ii) a phase with orientational order and complete translational order 
(iii) a smectic like phase in which orientational order is present and 
translational symmetry parallel to the orientation is broken.

To distinguish among the four different phases we define the following 
order parameters:
\begin{subequations}
\label{eq:order-parameter}
\bea
Q_1 &=& m^2 k \frac{\langle N_h-N_v \rangle}{N}, 
\label{eq:q1}\\
Q_2 &=& m^2 k\frac{\langle |\sum^{m^2-1}_{j=0} n_j e^{\frac{2\pi i j}{m^2}} | \rangle}{N}, 
\label{eq:q2}\\
Q_3 &=& m^2 k\frac{\langle |\sum_{j=0}^{m-1} r_j e^{\frac{2\pi i j}{m}} | 
-|\sum_{j=0}^{m-1} c_j e^{\frac{2\pi i j}{m}} |\rangle}{N}, \label{eq:q3}\\
Q_4&=& m^2 k \frac{\langle n_0-n_1-n_2+n_3 \rangle}{N}, \label{eq:q4}
\eea
\end{subequations}
where $N_h$ and $N_v$ are the total number of horizontal and vertical 
rectangles respectively, $n_i$ is the number of rectangles whose heads 
are in sublattice $i=0, \ldots, m^2-1$, $r_j$ are the number of 
rectangles whose heads are in row $(j\mod m)$, and $c_j$ are the 
number of rectangles whose heads are in column $(j \mod m)$. All 
four order parameters are zero in the I phase. $Q_1$ is non-zero in the 
N and C phases, $Q_2$ is non-zero in the C and S phases, $Q_3$ is 
non-zero only in the C phase, and $Q_4$ is non-zero only in the S phase. 
$Q_4$ in Eq.~(\ref{eq:q4}) has been defined for $m=2$.  Its 
generalization to $m \geq 3$ is straightforward.

We now define the thermodynamic quantities that are useful to 
characterize the transitions between the different phases.  $Q_i$'s 
second moment $\chi_i$, compressibility $\kappa$ and the Binder cumulant 
$U_i$ are defined as
\begin{subequations}
\label{eq:thermo-definition}
\bea
\chi_i&=&\langle Q_i^2 \rangle L^2, \label{eq:chi}\\
\kappa &= &[\langle \rho^2 \rangle-\langle \rho \rangle ^2] L^2,
\label{eq:kappa}\\
U_i&=&1- \frac{\langle Q_i^4 \rangle} {3 \langle Q_i^2 \rangle ^2}.
\label{eq:U}
\eea
\end{subequations}

The transitions are accompanied by the singular behavior of the above 
thermodynamic quantities at the corresponding critical densities. Let 
$\epsilon=(\mu-\mu_c)/\mu_c$, where $\mu_c$ is the critical chemical 
potential. The singular behavior is characterized by the critical 
exponents $\alpha$, $\beta$, $\gamma$, $\nu$ defined by $Q \sim 
(-\epsilon)^\beta$, $\epsilon<0$, $\chi \sim
|\epsilon|^{-\gamma}$, $\kappa \sim |\epsilon|^{-\alpha}$, and $\xi
\sim |\epsilon|^{-\nu}$, where $\xi$ is the correlation length, 
$|\epsilon| \rightarrow 0$, and $Q$ represents any of the order 
parameters. Only two exponents are independent, others being related to 
them through scaling relations.

The critical exponents $\alpha$, $\beta$, $\gamma$ and $\nu$ are 
obtained by finite size scaling of the different quantities near the 
critical point:
\begin{subequations}
\label{eq:scaling}
\bea
U &\simeq& f_u(\epsilon L^{1/\nu}), \label{eq:Uscaling}\\
Q &\simeq& L^{-\beta/\nu} f_q(\epsilon L^{1/\nu}), \label{eq:Qscaling}\\
\chi & \simeq& L^{\gamma/\nu} f_{\chi}(\epsilon L^{1/\nu}),
\label{eq:chiscaling}\\
\kappa & \simeq & L^{\alpha/\nu} f_\kappa(\epsilon L^{1/\nu}),
\label{eq:kappascaling}
\eea
\end{subequations}
where $f_u$, $f_q$, $f_{\chi}$, and $f_\kappa$ are scaling functions.

\section{\label{sec:phase_diag2}Phase Diagram and Critical Behavior for $m=2$}

In this section, we discuss the phase diagram for the case $m=2$ and 
aspect ratio $k=1,2,\ldots 7$. The critical exponents characterizing the 
different continuous transitions are determined numerically.

\subsection{Phase Diagram}

\begin{figure}
\includegraphics[width=\columnwidth]{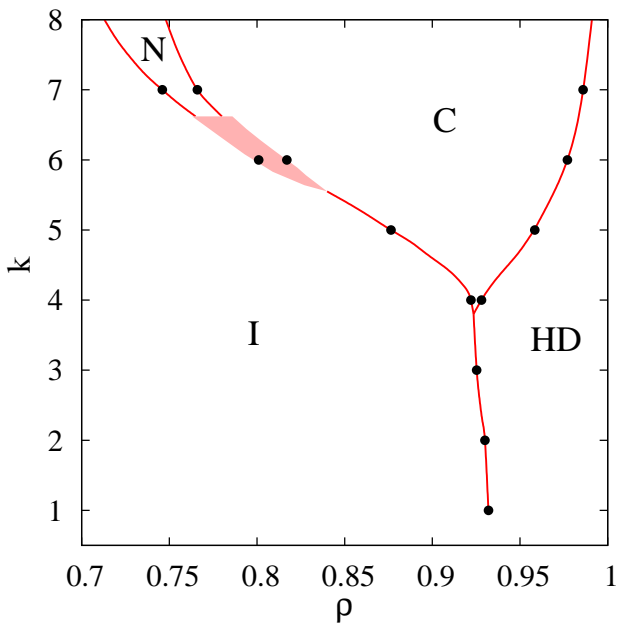}
\caption{(Color online) Phase diagram for rectangles of size $2 \times 2 
k$. I, N, C and HD denote isotropic, nematic, columnar and high density 
phases respectively. The HD phase is a C phase for $k=1$ and a S phase 
for $k>1$. The data points are from simulation, while the continuous 
lines and shaded portions are guides to the eye. 
The shaded portion denotes regions of phase 
coexistence. }
\label{fig:phase_diag_m2} 
\end{figure}

The phase diagram obtained from simulations for $m=2$ and integer $k$ 
are shown in Fig.~\ref{fig:phase_diag_m2}. The low density phase is an I 
phase for all $k$. The case $k=1$ is different from other $k$. In this 
case, the problem reduces to a hard square problem and orientational 
order is not possible as there is no distinction between horizontal and 
vertical rectangles. The hard square system undergoes only one 
transition with increasing density, it being a continuous transition 
from the I phase to a C 
phase~\cite{baxterBook,nigam1967,pearce1988,kabir2012}. This transition 
belongs to the Ashkin Teller universality class (see 
Refs.~\cite{fernandes2007,hirokazu2007, nienhuis2011,kabirthesis} for 
recent numerical studies). For $k =2, 3$, we find that the system 
undergoes one continuous transition directly from the I phase to a 
crystalline S phase.  On the other hand, the system with $k=4, 5, 6$ may 
exist in I, C, or S phases. With increasing density, the system 
undergoes two phase transitions: first from the I to a C phase which 
could be continuous or first order, and second, from the C to a S phase 
which is continuous.  For $k=7$, we observe three continuous transitions 
with increasing density: first from the I to the N phase, second into 
the C phase and third into the S phase. By confirming the existence of 
the N and C phases for $k=8$, we expect the phase behavior for $k\geq 8$ 
to be similar to that for $k=7$.

\begin{figure}
\includegraphics[width=\columnwidth]{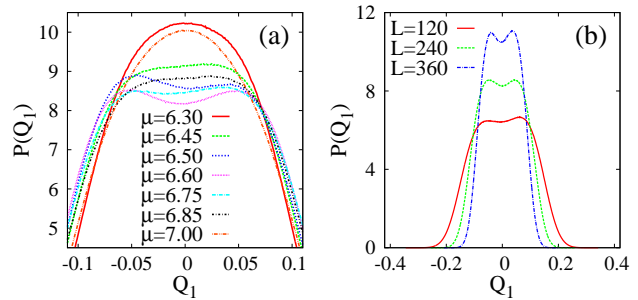}
\caption{(Color online) (a) The probability distribution of the order 
parameter $Q_1$ for $k=4$ at different $\mu$ values, when $L=416$. (b) 
The same for different system sizes when $\mu=6.55$.}
\label{fig:w2k8}
\end{figure}

The system undergoes more than one transition only for $k \geq 
k_{min}=4$. We now present some supporting evidence for this claim. In 
Fig.~\ref{fig:w2k8} (a) we show the probability distribution of the 
nematic order parameter $Q_1$, when $k=4$, for different values of $\mu$ 
and fixed $L$, close to the I-C and the C-S transitions. For lower values of 
$\mu$, the distribution is peaked around zero corresponding to the I 
phase. With increasing $\mu$, the distribution becomes flat and two 
symmetric maxima appear at $Q_1 \neq 0$ ($Q_3$ also becomes nonzero 
simultaneously), corresponding to a C phase. On increasing $\mu$ 
further, the two maxima continuously merge into a single peak at 
$Q_1$=0, corresponding to the S phase ($Q_3$ also becomes zero and $Q_4$ 
becomes nonzero). Fig.~\ref{fig:w2k8} (b) shows the distribution of 
$Q_1$ for three different system sizes at a fixed value of $\mu$ for 
which $P(Q_1)$ has two symmetric maxima at $Q_1 \neq 0$. The two peaks 
become sharper and narrower with increasing $L$. We find the similar 
behavior for $P(Q_3)$ also. From the above, we conclude that the C phase 
exists for $k=4$ albeit for a very narrow range of $\mu$. For $k=2$ and 
$3$ we do not observe the existence of a columnar phase and find that 
the probability distributions of $Q_1$ and $Q_3$ are peaked around zero 
for all $\mu$. Hence, we conclude that $k_{min}=4$.

The N phase exists only for $k\geq 7$. This is also true for 
$m=1$~\cite{ghosh2007}. To see this, notice that the I-C transition 
for $k=6$ is 
first order (see Fig.~\ref{fig:phase_diag_m2}). If a nematic phase 
exists for $k =6$, then the first transition would have been continuous 
and in the Ising universality class~\cite{fernandez2008a}.

\subsection{Critical behavior for the isotropic-sublattice (I-S) phase 
transition}

\begin{figure}
\includegraphics[width=\columnwidth]{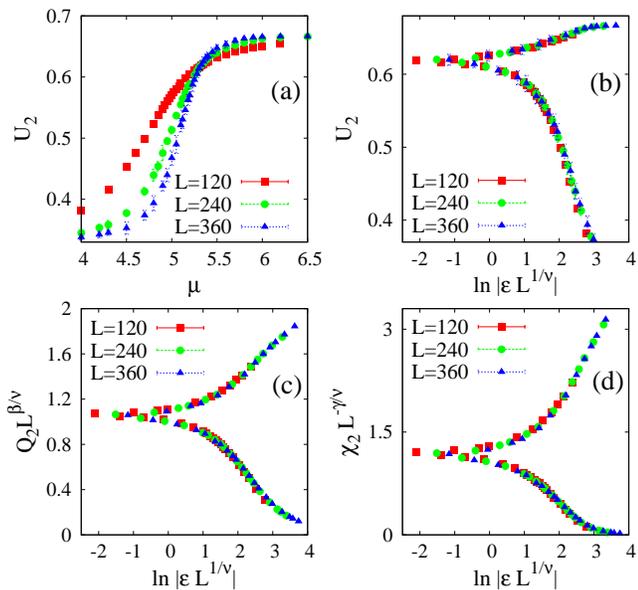}
\caption{(Color online) Data collapse for different $L$ near the I-S 
transition for rectangles of size $2\times 4$ ($m=2$, $k=2$). We find 
$\mu_c \approx 5.33$ ($\rho_c \approx 0.93$). The exponents are 
$\beta/\nu=1/8$, $\gamma/\nu=7/4$ and $\nu \approx 1.18$.}
\label{fig:scaling_K4}
\end{figure}

\begin{figure}
\includegraphics[width=\columnwidth]{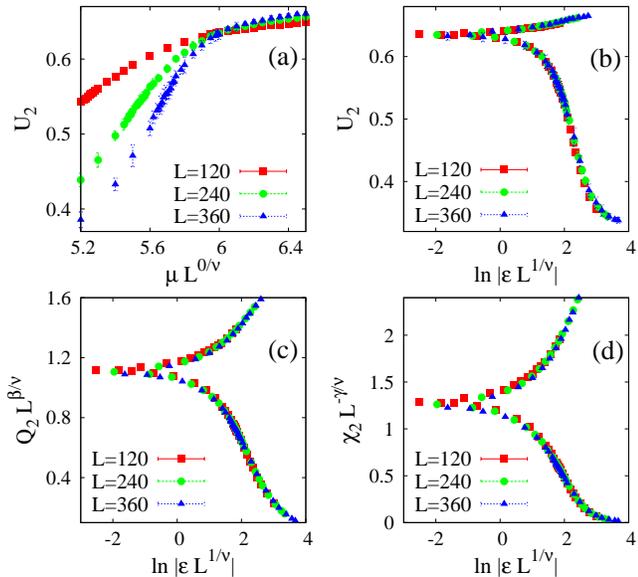}
\caption{(Color online) Data collapse for different $L$ near the I-S 
transition for rectangles of size $2\times 6$ ($m=2$, $k=3$). We find 
$\mu_c \approx 6.04$ ($\rho_c \approx 0.925$). The exponents are 
$\beta/\nu=1/8$, $\gamma/\nu=7/4$ and $\nu \approx 1.23$}.
\label{fig:scaling_K6}
\end{figure}

The system of rectangles with $m=2$ undergoes a direct I-S transition 
for $k=2,3$. At this transition, the translational symmetry gets broken 
along both $x$- and $y$- direction but the rotational symmetry remains 
preserved. 
We study this transition using the order parameter $Q_2$ 
[see Eq.~(\ref{eq:q2})]. $Q_2$ is non-zero in the S phase and zero in 
the I phase. In this case $Q_1$ and $Q_3$ remains zero 
for all values of $\mu$. 
The data collapse of $U_2$, $Q_2$, and $\chi_2$ for 
different values of $L$ near the I-S transition are shown in 
Fig.~\ref{fig:scaling_K4} for $k=2$ and in Fig.~\ref{fig:scaling_K6} for 
$k=3$. From the crossing of the Binder cumulant data for different $L$, 
we estimate the critical chemical potential $\mu_c \approx 5.33$ 
($\rho_c \approx 0.930$) for $k=2$ and $\mu_c \approx 6.04$ ($\rho_c 
\approx 0.925$) for $k=3$. The order parameter increases continuously 
with $\mu$ from zero as $\mu_c$ is crossed, making the transition 
continuous.  Since the $S$ phase has a four fold symmetry due to the 
four possible sublattices, we expect the 
transition to be in the Ashkin-Teller universality class. Indeed, we 
find a good collapse with $\beta/\nu=1/8$ and $\gamma/\nu=7/4$. 
Numerically, we find $\nu=1.18\pm0.06$ for $k=2$ and $\nu=1.23\pm 0.07$ 
for $k=3$. $\nu$ being larger than $1$, we do not observe any divergence 
in $\kappa$. This transition could have also been studied using the order parameter $Q_4$.

\subsection{Critical behavior of the isotropic-columnar phase (I-C) transition}

The I-C transition is seen for $k=4,5,6$. When for $k=4$, $\mu_c$ for
the I-C and the C-S transitions are close to each other, making
$k=4$ unsuitable for studying the critical behavior. We, therefore,
study the I-C transition for $k=5$ ($2 \times 10$ rectangles) and $k=6$
($2 \times 12$ rectangles). 

\begin{figure}
\includegraphics[width=\columnwidth]{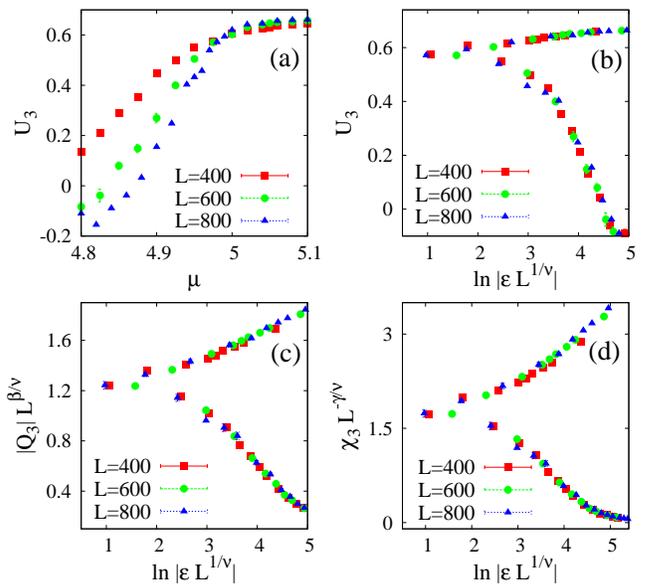}
\caption{(Color online) The data 
for different $L$ near the I-C transition collapse  when scaled with exponents
$\beta/\nu=1/8$, $\gamma/\nu=7/4$, and $\nu = 0.82$. We find
$\mu_c \approx 4.98$ ($\rho_c$ $\approx 0.876$). 
Data are for rectangles of size $2 \times 10$.}
\label{fig:exponents_K10f}
\end{figure}

The critical behavior is best studied using the order parameter $Q_3$
[see Eq.~(\ref{eq:q3})]. $Q_3$ is non-zero only in the C phase. 
First, we present the critical behavior for $k=5$.
The simulation data for different system sizes are shown in
Fig.~\ref{fig:exponents_K10f}. From the crossing of the Binder
cumulant curves, we obtain $\mu_c \approx 4.98$ ($\rho_c \approx
0.876$). The transition is found to be continuous. There are four possible
columnar states:  majority of  heads are either in even or odd rows 
(when horizontal orientation is preferred), or in 
even or odd columns 
(when vertical orientation is preferred). Due to this four fold symmetry,
we expect the I-C transition to be in the 
Ashkin-Teller universality class.
The data for different $L$ collapse with $\beta/\nu=1/8$,
$\gamma/\nu=7/4$ and $\nu = 0.82 \pm 0.06$ (see Fig.~\ref{fig:exponents_K10f}),
confirming the same.
Unlike the I-S transition for $k=2,3$, $\nu<1$ and lies between the 
Ising and q=4 Potts points. At the I-C transition partial breaking of 
translational symmetry and complete breaking of rotational symmetry 
occur simultaneously. 

The I-C transition for $k=4$ is also continuous [see Fig.~\ref{fig:w2k8} (a)], 
and is therefore  expected to be in the Ashkin-Teller universality class.
However, there is no reason to
expect that $\nu$ will be the same as that for $k=5$. 

\begin{figure}
\includegraphics[width=\columnwidth]{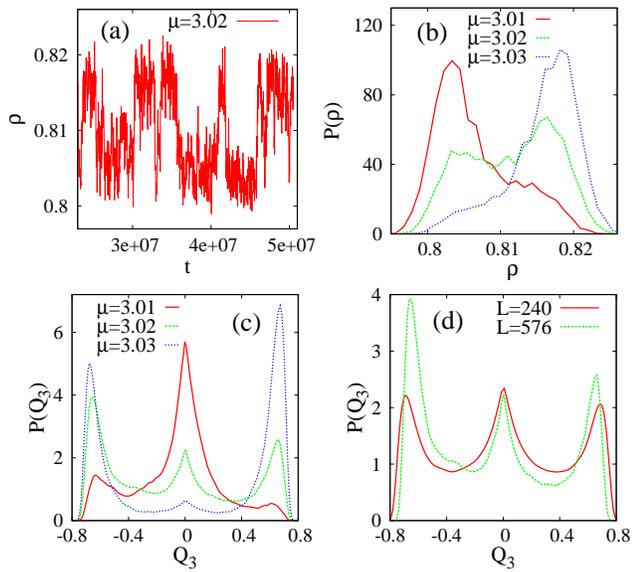}
\caption{(Color online) Data  for the density $\rho$ and 
the order parameter
$Q_3$ for rectangles of size $2\times 12$. (a) Equilibrium time profile of 
$\rho$ near the I-C transition, for $\mu=3.02$ and $L=720$.
Probability distribution, near the I-C transition, of 
(b) $\rho$ for different
values of $\mu$ when $L=576$,
(c) $Q_3$ for different
values of $\mu$ when $L=576$, and (d) $Q_3$ for 
$L=240$ ($\mu=2.99$) and $L=576$ ($\mu=3.02$).
}
\label{fig:dist_K12f}
\end{figure}

For $k=6$, the I-C transition is surprisingly first order.
Fig.~\ref{fig:dist_K12f}(a) shows the time profile of density near
the I-C transition. $\rho$ alternates between two well defined
densities, one corresponding to the I phase and the other to the C
phase. This is also seen in the probability distribution for density
[see Fig.~\ref{fig:dist_K12f}(b)].
Near the I-C transition it shows 
two peaks corresponding to the I and the C phases. 
Thus, at $\mu=\mu_c^{IC}$, the density has a discontinuity, which is
shown by the shaded region in the phase diagram 
(see Fig.~\ref{fig:phase_diag_m2}). 
The probability distribution of the order parameter 
$Q_3$ shows similar behavior [see Fig.~\ref{fig:dist_K12f}(c)]. 
Near $\mu=\mu_c^{IC}$ 
the distribution shows three peaks: one at $Q_3=0$ corresponding to the I phase and 
the other two at $Q_3\neq 0$, corresponding to the C phase. At 
$\mu_c^{IC}$ the three peaks become of equal height and the order parameter 
$Q_3$ 
jumps from zero to a non-zero value. These peaks sharpen with
increasing system size [see Fig.~\ref{fig:dist_K12f}(d)]. These are 
typical signatures of a first order transition. 
Hence, we conclude that the I-C transition 
may be continuous or first order depending on $k$. 

\subsection{Critical behavior of the isotropic-nematic phase (I-N) transition}

\begin{figure}
\includegraphics[width=\columnwidth]{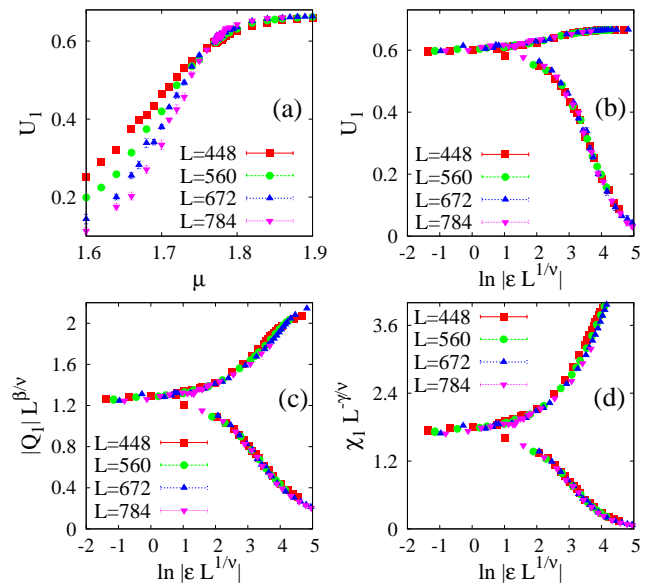}
\caption{(Color online) 
The data for different $L$ 
near the I-N transition collapse when scaled with the
Ising exponents $\beta/\nu=1/8$, $\gamma/\nu=7/4$, $\nu = 1 $ and $\mu_c \approx 1.77$. 
The critical density $\rho_c$ $\approx 0.746$. Data are for rectangles of size $2 \times 14$.}
\label{fig:exponents_K14IN}
\end{figure}

We find that the nematic phase exists only for $k \geq 7$. We study the I-N phase transition
for $k=7$ using the order parameter $Q_1$ [see Eq.~(\ref{eq:q1})]. $Q_1$ is non-zero in the 
N and C phases and zero in the I phase. 
We confirm that the ordered phase is an N phase by checking that
$Q_3$, which is non-zero only in the C phase, is zero. 
In the nematic phase the rectangles may choose either horizontal 
or vertical orientation. Thus,
we expect the transition to be in the Ising universality class. When $m=1$, this
has been verified using extensive Monte Carlo simulations~\cite{fernandez2008a}. Here,
we confirm the same for $m=2$. The 
data for $U_1$, $|Q_1|$ and $\chi_1$ for different $L$ collapse onto one
curve when scaled with the two dimensional Ising exponents
$\beta/\nu=1/8$, $\gamma/\nu = 7/4$, and $\nu=1$ (see Fig.~\ref{fig:exponents_K14IN}).
We find $\mu_c\approx1.77$ ($\rho_c \approx 0.746$). We note that the value of
$U_2$ at the point where the curves for different $L$ cross is slightly smaller than
the Ising value $0.614$. This suggests that larger system sizes are necessary for better 
collapse of the data.

\subsection{Critical behavior of the nematic-columnar phase (N-C) transition}

\begin{figure}
\includegraphics[width=\columnwidth]{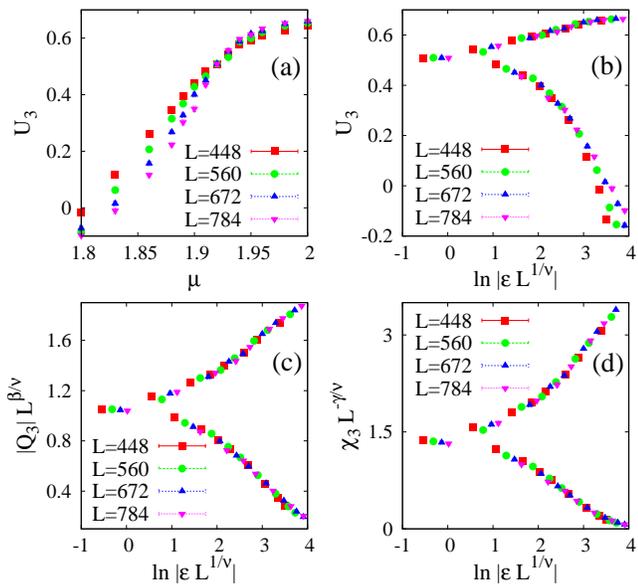}
\caption{(Color online) 
The data for different $L$ 
near the N-C transition collapse  when scaled with the
Ising exponents
$\beta/\nu=1/8$, $\gamma/\nu=7/4$, $\nu = 1 $ and $\mu_c \approx 1.92$. The critical 
density $\rho_c$ 
$\approx 0.766$. Data are for rectangles of size $2 \times 14$.}
\label{fig:exponents_K14NC}
\end{figure}

The N-C transition is also studied for $k=7$, using the order parameter $Q_3$. $Q_3$ 
is zero in the nematic phase but nonzero in the columnar phase. At
the I-N transition 
orientational symmetry gets broken. 
If the nematic phase 
consists of mostly horizontal (vertical) rectangles, then 
there is no preference over even and odd rows (columns). In the columnar phase 
the system chooses either even or odd rows (columns), once the orientational symmetry is broken. 
Due to the two broken symmetry phases we expect this transition to be in the 
Ising universality class. We indeed find good data collapse 
when $U_3$, $|Q_3|$ and $\chi_3$ 
for different system sizes are scaled 
with Ising exponents (see Fig.~\ref{fig:exponents_K14NC}). 
The critical chemical potential or critical density is obtained from 
the crossing point of the binder 
cumulant $U_3$ for different $L$. 
We find $\mu_c \approx 1.92$ ($\rho_c \approx 0.766$) for this transition. 
We expect the critical behavior to be same for $k >7$. 

\subsection{Critical behavior of the columnar-sublattice phase (C-S) transition}

\begin{figure}
\includegraphics[width=\columnwidth]{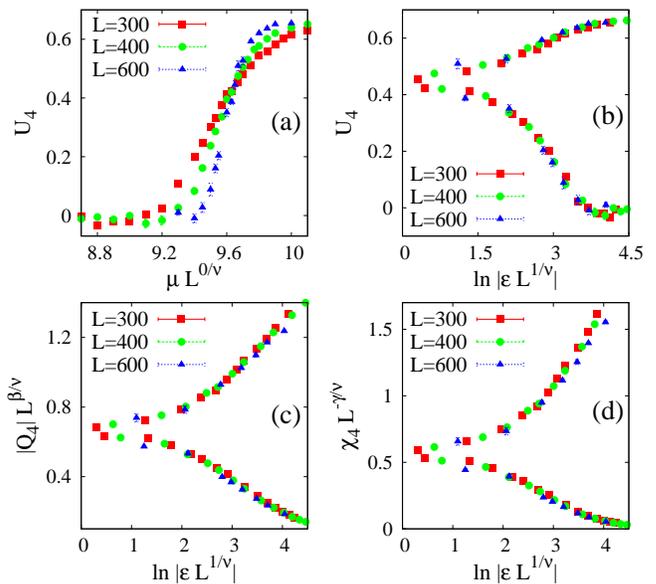}
\caption{(Color online) The data 
for different $L$ near the C-S transition collapse  when scaled with exponents
$\beta/\nu=1/8$, $\gamma/\nu=7/4$, $\nu = 0.83$ and $\mu_c \approx 9.65$. The critical 
density $\rho_c$ $\approx 0.958$. Data are for rectangles of size $2 \times 10$.}
\label{fig:exponents_K10s}
\end{figure}

The C-S transition exists for $k\geq 4$. 
This transitions is studied by choosing 
$k=5$. We characterize the C-S transition using the order
parameter $Q_4$ which is non zero only in the S phase. In the C phase 
the system chooses one particular orientation and either even or 
odd rows or columns, depending on the orientation. This corresponds to 
two sublattices being chosen among four of them. In the C-S 
transition the translational symmetry gets broken completely by choosing a 
particular sublattice, but along with that the orientational symmetry gets 
restored. This transition is found to be continuous. The data of $U_4$, 
$|Q_4|$ and 
$\chi_4$ for different $L$ near the C-S transition collapse
well when scaled with the exponents belonging to the Ashkin-Teller universality class. The estimated 
critical exponents are $\beta/\nu=1/8$,
$\gamma/\nu=7/4$ and $\nu = 0.83 \pm 0.06$ (see Fig.~\ref{fig:exponents_K10s}). Binder 
cumulants for different system sizes cross at $\mu_c \approx 9.65$ ($\rho_c \approx 0.958$). We
expect similar behavior for $k=4$ and $6$ but possibly with different $\nu$. 
The C-S transition occurs at very high density. With increasing $k$, 
the relaxation time becomes increasingly large, making it difficult to
obtain reliable data for the C-S transition when $k \geq 6$.

\section{\label{sec:phase_diag3}Phase Diagram and Critical Behavior for $m=3$}

\subsection{Phase diagram}

\begin{figure}
\includegraphics[width=\columnwidth]{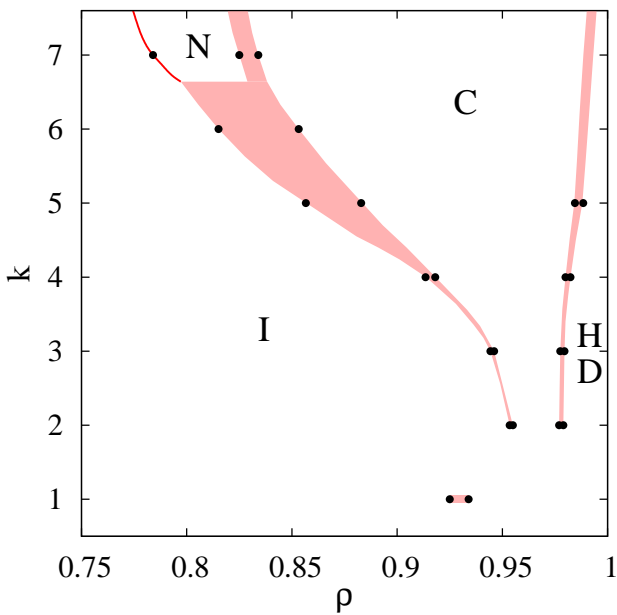}
\caption{(Color online) Phase diagram for rectangles of size $3 \times 3 
k$. HD denotes high density. The HD phase is a C phase for $k=1$ and a S 
phase for $k>1$. The data points are from simulation while the 
continuous line and shaded portions are a guide to the eye. 
The shaded portions denote 
regions of phase coexistence. Except the I-N transition, all the 
transitions are found to be first order.}
\label{fig:phase_diag_m3}
\end{figure}

The phase diagram that we obtain for $m=3$, is shown in 
Fig.~\ref{fig:phase_diag_m3}. When $k=1$, the corresponding hard square 
system has a single, first order transition from the I phase into the C 
phase~\cite{fernandes2007}. The shaded region between two points denotes 
a region of phase coexistence. For $2 \leq k \leq 6$, the system 
undergoes two first order transitions with increasing density: first an 
I-C transition and second a C-S transition. This is unlike the case 
$m=2$, where for $k=2$ and $3$ we find only one transition. For $k=7$, 
we find three transitions as in the $m=2$ case. The first transition 
from I to N phase is continuous while the second from N to C phase 
appears to be first order.  Although we cannot obtain reliable data for 
the third transition into the S phase, we expect it to be first order. 
We note that the minimum value of $k$ beyond which the nematic phase 
exists is $7$ for both $m=2$ and $m=3$, and matches with that for 
$m=1$~\cite{ghosh2007}.

\subsection{The isotropic-columnar phase (I-C) transition}

\begin{figure}
\includegraphics[width=\columnwidth]{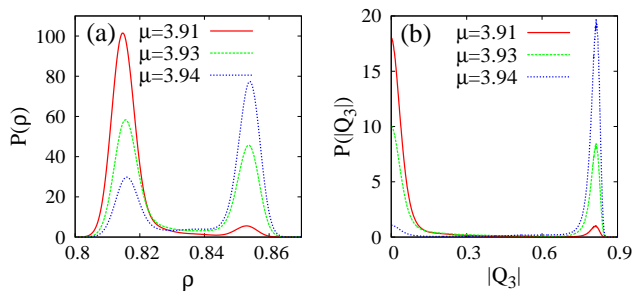}
\caption{(Color online) Distribution of (a) the density $\rho$ near the 
I-C transition, (b) the order parameter $|Q_3|$ near the I-C transition. 
The data are for rectangles of size $3\times 18$ and $L=432$.} 
\label{fig:w3k18f} 
\end{figure}

The I-C transition exists when $k \leq 6$. We study the this transition 
for $k=6$, using the order parameter $Q_3$. Now there are six possible 
choices for the C phase: heads are predominantly in one of the rows 
$0,1$ or $2$ (mod $3$) with all the columns equally occupied (if 
horizontal orientation is preferred) or in one of the columns $0,1$ or 
$2$ (mod $3$) and all the rows are equally occupied (if vertical 
orientation is preferred). Making an analogy with the six state Potts 
model, we expect the I-C transition to be first order. The probability 
distribution of the density $\rho$ and the order parameter $|Q_3|$ for 
$k=3$ near the I-C transition is shown in Fig.~\ref{fig:w3k18f}. The 
distributions are clearly double peaked at and near the transition 
point, one corresponding to the I phase and the other to the 
C phase. We find that these peaks become sharper with increasing 
system size. This is suggestive of a first order phase transition with a 
discontinuity in both density and order parameter as $\mu$ crosses 
$\mu_c$. The discontinuity in the density is denoted by the shaded 
regions of Fig.~\ref{fig:phase_diag_m3}. The chemical potential at which 
the I-C transition occurs is given by, $\mu^{\ast}_{IC} \approx 3.93$. 
Similar behavior is seen near the I-C transition for rectangles of size 
$3\times 3 k$ with $k=2,3,4$ and $5$. We observe that the discontinuity 
in the density increases with $k$.

\subsection{Critical behavior of the isotropic-nematic phase (I-N) transition}

\begin{figure}
\includegraphics[width=\columnwidth]{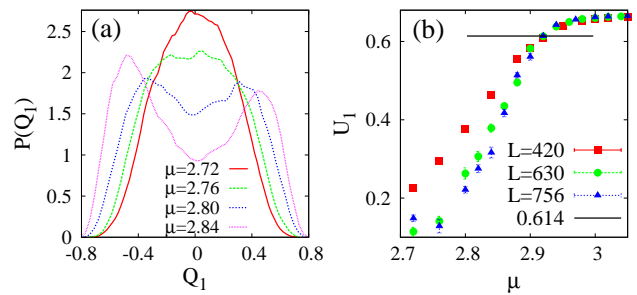}
\caption{(Color online) 
The I-N transition for rectangles of size $3\times 
21$.
(a) Distribution of the order parameter $Q_1$ 
near the I-N transition. The data are for  
$L=420$.
(b) Binder cumulant for different system sizes 
crosses at $\mu_c \approx 2.92$ ($\rho_c\approx 0.79$). Value of $U$ at 
$\mu_c$ is $\approx 0.61$. 
}
\label{fig:w3k21IN} 
\end{figure}

As for $m=2$, for $m=3$ we find the existence of the nematic phase only 
for $k\geq 7$. We study the I-N transition for $k=7$ with the order 
parameter $Q_1$. It is expected to be in the Ising universality class 
since there are two possible choices of the orientation: either 
horizontal or vertical. 
We are unable to obtain good data collapse for $|Q_1|$, $\chi_1$ and $U_1$ 
as the relaxation time increases with increasing $m$ and $k$. Instead,
we present some evidence for the transition being continuous and
belonging to the Ising universality class.
In Fig.~\ref{fig:w3k21IN} (a), the distribution of the order 
parameter $Q_1$ near the I-N transition is shown. 
The two symmetric peaks of the 
distribution come closer with decreasing $\mu$ and merge 
to a single peak, this being a signature of a continuous 
transition. 
The Binder cumulant $U_1$ for different system 
sizes cross at $\mu_c \approx 2.92$ ($\rho_c\approx 0.79$) [see 
Fig.~\ref{fig:w3k21IN} (b)]. The value of $U$ at $\mu=\mu_c^{I-N}$ is 
very close to the $U_c$ value ($0.61$) for the Ising universality class. 

\subsection{The nematic-columnar phase (N-C) transition}

\begin{figure}
\includegraphics[width=\columnwidth]{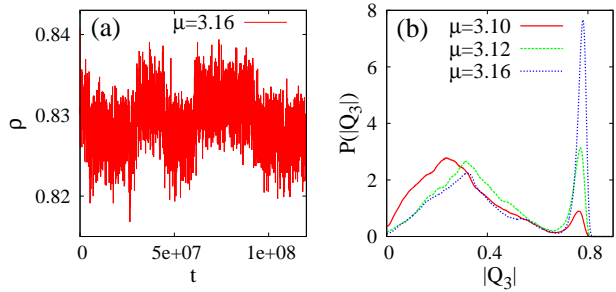}
\caption{(Color online) (a) Temporal variation of density near the N-C 
transition, (b) distribution of the order parameter $|Q_3|$ near the N-C 
transition. The data are for rectangles of size $3\times 21$ and 
$L=756$.} 
\label{fig:w3k21NC} 
\end{figure}

The N-C transition is studied for $k=7$ using the order parameter 
$Q_3$. Contrary to our expectation that the N-C transition should be 
in the q=3 Potts universality class, we observe a first order 
transition. The temporal dependence of the density near the N-C 
transition is shown in Fig.~\ref{fig:w3k21NC}(a). Density jumps between 
two well separated values corresponding to the two different phases near 
the coexistence. Fig~\ref{fig:w3k21NC}(b) shows the discontinuity in the 
order parameter $|Q_3|$ near the transition. $P(|Q_3|)$ shows two peaks 
of approximately equal height near $\mu{^\ast}_{N-C} \approx 3.12$. 
However, we are limited in our ability to obtain reliable data for 
$3\times 21$ rectangles for larger system sizes, and the observed first 
order nature could be spurious.

\subsection{The columnar-sublattice phase (C-S) transition} 

\begin{figure}
\includegraphics[width=\columnwidth]{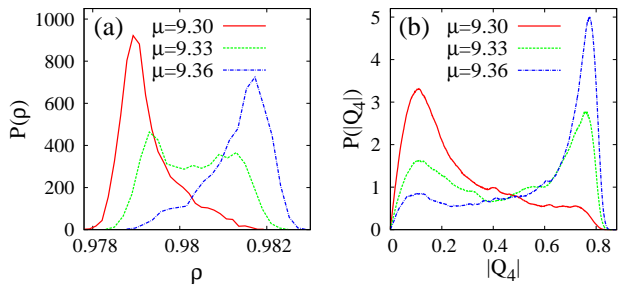}
\caption{(Color online) Distribution of (a) the density $\rho$ near the 
C-S transition and (e) the order parameter $|Q_4|$ near the C-S 
transition. The data are for rectangles of size $3\times 6$ and 
$L=720$.} 
\label{fig:w3k6s} 
\end{figure}

The C-S transition is studied by choosing $k=2$. We use the order 
parameter $Q_4$ which is nonzero only in the S phase. The probability 
distribution of the density $\rho$ and the order parameter $|Q_4|$ for 
$3\times 6$ rectangles near the C-S transitions is shown in 
Fig.~\ref{fig:w3k6s}. The distributions are again double peaked at and 
near the transition point, making the C-S transition first order. These 
peaks become sharper with increasing system size. The discontinuity in 
the density near the C-S transition is very small and can also be seen 
in the shaded portions of Fig.~\ref{fig:phase_diag_m3}. We estimate 
$\mu^{\ast}_{C-S} \approx 9.33$. Similar behavior near the C-S 
transitions is also observed for $k > 2$, but the relaxation 
time increases with $k$. 

\section{\label{sec:phase_boundary}Estimation of the phase boundaries 
using analytical methods}

In this section we obtain the asymptotic behavior of the phase diagram 
for large $k$ using theoretical arguments.

\subsection{The isotropic--nematic phase boundary}

The critical density for the I-N phase transition, for fixed $m$ and $k 
\gg 1$ may be determined by making an analogy with the continuum 
problem. The limit $k\to \infty$, keeping $m$ fixed corresponds to the 
system of oriented lines in the continuum. For this problem $\rho_c 
\approx A_1/k$~\cite{ghosh2007,fernandez2008c}. The constant $A_1$ is 
estimated to be $\approx 6.0$~\cite{ghosh2007,fernandez2008c}. Thus, we 
expect $\rho_c^{I-N} \approx A_1/k$, where $A_1$ is independent of $m$. 
For $k=7$, we observe only a weak dependence of $\rho_c^{I-N}$ on $m$ 
with the critical density being $0.745\pm 0.005$ $(m=1)$, $0.744 \pm 
0.008$ $(m=2)$ and $0.787 \pm 0.010$ $(m=3)$.

\subsection{The nematic--columnar phase boundary}

To obtain the asymptotic behavior of the N-C phase boundary, we use an 
ad hoc Bethe approximation scheme for rods due to 
DiMarzio~\cite{dimarzio1961}, adapted to other 
shapes~\cite{sokolova2000}. To estimate the phase boundary of the 
nematic--columnar transition of $m\times m k$ rectangles on the square 
lattice with $M=L\times L$ sites, we require the entropy as a function 
of the occupation densities of the $m$ types of rows/columns. The 
calculations become much simpler, if we consider a fully oriented phase 
with only horizontal rectangles. Now, the nematic phase corresponds to 
the the phase where there is equal occupancy of each of the $m$ types of 
rows, while the columnar phase breaks this symmetry and preferentially 
occupies one type of row. For this simplified model with only one 
orientation, we estimate the entropy within an ad hoc Bethe 
approximation as detailed below. We present the calculation for $m=2$, 
classifying the rows as even and odd rows. Generalization to higher 
values of $m$ is straight forward.

\begin{figure}
\includegraphics[width=\columnwidth]{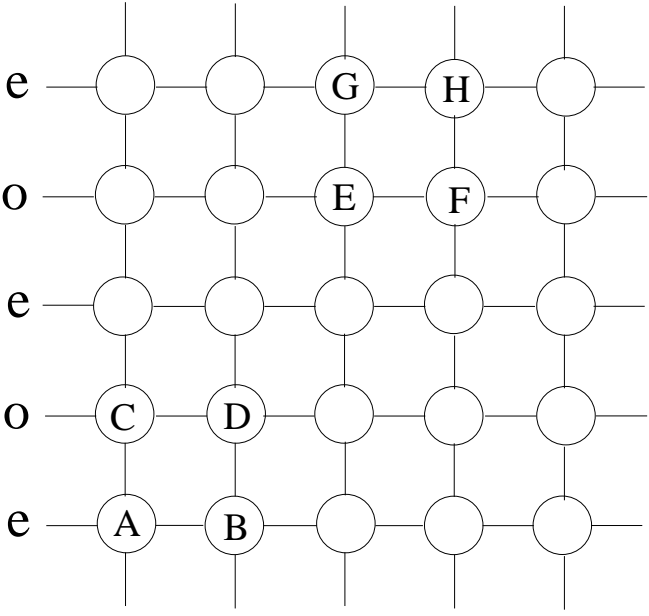}
\caption{(Color online) Schematic diagram showing the positions of
sites $A$--$F$ to aid the explanation of the calculation of the
nematic--columnar phase boundary.
Even and odd
rows are denoted by e and o respectively.}
\label{fig:lat_nc}
\end{figure}

Let there be $N_e$ ($N_o$) number of rectangles whose heads (left bottom 
site of the rectangle) occupy even (odd) rows. We first place the $N_e$ 
rectangles one by one on the even rows. Given that $j_e$ rectangles have 
already been placed, the number of ways in which the $(j_e+1)^{th}$ 
rectangle can be placed may be estimated as follows. The head of the 
$(j_e+1)^{th}$ rectangle has to be placed on an empty site of an even 
row. We denote this site by $A$ (see Fig.~\ref{fig:lat_nc}). The site 
$A$ can be chosen at random in ($M/2-2 k j_e$) ways, $M/2$ being the 
number of sites in even rows and $2 k j_e$ being the number of occupied 
sites in the even rows by the $j_e$ rectangles. We now require that the 
$2 k -1$ consecutive sites to the right of $A$ are also empty. The 
probability of this being true is $[P_x(B|A)]^{2 k -1}$, where 
$P_x(B|A)$ is the conditional probability that $B$ (see 
Fig.~\ref{fig:lat_nc}) is empty given that $A$ is empty. In terms of $M$ 
and $j_e$, $P_x(B|A)$ is given by
\be
P_x(B|A)=\frac{\frac{M}{2}-2 k j_e}{\frac{M}{2}-2 k j_e+j_e}.
\ee

To place the $(j_e+1)^{th}$ rectangle, we also require that the site $C$ 
(see Fig.~\ref{fig:lat_nc}) and the $2 k -1$ consecutive sites to the 
right of $C$ are also empty. The probability of this being true is 
$P_y(C|A) \times P_{xy}(D|B \cap C)^{2 k -1}$, where $P_y(C|A)$ is the 
conditional probability that $C$ is empty given $A$ is empty, and 
$P_{xy}(D|B \cap C)$ is the conditional probability that $D$ (see 
Fig.~\ref{fig:lat_nc}) is empty given that both $B$ and $C$ are empty.  
Sites C and D belongs to an odd row. Since both A and B are empty, C and 
D can be occupied only by rectangles with heads in the same odd row. 
But, there are no such rectangles. Therefore, $P_y(C|A)=1$, and 
$P_{xy}(D|B \cap C)=1$. Thus, given that $j_e$ rectangles have been 
placed, the $(j_e+1)^{th}$ rectangle may be placed in
\be
\nu_{j_e+1} = \left(\frac{M}{2}-2 k j_e \right ) \times [P_x(B|A)]^{2k-1} 
\ee
ways.
Hence, the total number of ways of placing $N_e$ rectangles with heads
on even rows is 
\bea
\Omega_e&=&\frac{1}{N_e!} \displaystyle\prod_{j_e=0}^{N_e-1} \nu_{j_e+1} \nonumber\\
&=& \frac{1}{N_e!} \prod_{j_e=0}^{N_e-1} \frac{(\frac{M}{2}-2 k j_e)^{2k}}
{(\frac{M}{2}-2 k j_e+j_e)^{2k-1}}.
\label{eq:omegaeven}
\eea

Keeping the $N_e$ rectangles with heads on even rows, we now place $N_o$ 
rectangles one by one on the odd rows. Given that $j_o$ rectangles have 
been placed on the odd rows, the number of ways of placing the 
$(j_o+1)^{th}$ rectangle may be estimated as follows. The head of the 
$(j_o+1)^{th}$ rectangle must be placed on an empty site on an odd row. 
We denote this site by $E$ (see Fig.~\ref{fig:lat_nc}). $E$ may be 
chosen in $(\frac{M}{2}-2kN_e-2 k j_o)$ ways, where we have ignored 
correlations between rectangles. Here $2kN_e$ is the number of occupied 
sites in the odd rows due to the $N_e$ rectangles on the even rows, and 
the $2 k j_o$ is the number of sites occupied by $j_o$ rectangles in odd 
rows. We now require that the $2 k -1$ consecutive sites to the right of 
$E$ are also empty. The probability of this being true is $[P_x(F|E)]^{2 
k -1}$, where $P_x(F|E)$ is the conditional probability that $F$ (see 
Fig.~\ref{fig:lat_nc}) is empty given that $E$ is empty. $P_x(F|E)$ is 
given by
\be
P_x(F|E)=\frac{M/2-2kN_e-2kj_o}{M/2-2k N_e-2kj_o+N_e+j_o},
\label{eq:7}
\ee 
where we have again ignored all correlations.

For placing the $(j_o+1)^{th}$ rectangle, we also require that the
site $G$ (see Fig.~\ref{fig:lat_nc}) and the $2 k -1$
consecutive sites to the right of $G$ are also empty. The probability
of this being true is $P_y(G|E) \times P_{xy}(H|F \cap G)^{2 k -1}$,
where $P_y(G|E)$ is the conditional probability that $G$ is empty
given $E$ is empty, and $P_{xy}(H|F \cap G)$ is the conditional
probability that $H$ (see Fig.~\ref{fig:lat_nc}) is empty given that
both $F$ and $G$ are empty. Ignoring correlations,
$P_y(G|E)$ is given by
\be
P_y(G|E)=\frac{M/2-2kN_e-2kj_o}{M/2-2kj_o}.
\label{eq:8}
\ee 
If we calculate $P(H|F \cap G)$ following the procedure 
developed by DiMarzio in
Ref.~\cite{dimarzio1961}, then the resultant entropy is not symmetric with respect
to $N_e$ and $N_o$ and depends on the order of placement.
To overcome this shortcoming, we follow the Bethe approximation
proposed in Ref.~\cite{sokolova2000} as follows:
\bea
P(H|F \cap G) &=& \frac{P(F \cap G |H) P(H)}{P_{xy}(G|F) P(F)},
\label{eq:9}\\
&\approx& \frac{P_x(G|H) P_y(F|H)}{P_{xy}(G|F)}.
\label{eq:10}
\eea
where in Eq.~(\ref{eq:9}), we used $P(H)=P(F)$ and in Eq.~(\ref{eq:10}), we
replaced $P(F \cap G |H) P(H)$ by $P_x(G|H) P_y(F|H)$, which is an
approximation.

In Eq.(\ref{eq:10}), from symmetry, it is easy to see that  
$P_x(G|H)=P_x(F|E)$ and $P_y(F|H)=P_y(G|E)$ and can be read off from
Eqs.~(\ref{eq:7}) and (\ref{eq:8}). To obtain an expression for $P_{xy}(G|F)$,
the probability that $G$ is empty, given that the site $F$ is empty,
we again ignore correlations. We then obtain
\be
P_{xy}(G|F)=\frac{M/2-2 k N_e-2 k j_o}{M/2-2 k j_o+j_o}.
\ee

The number of ways 
of placing the $(j_o+1)^{th}$ rectangle is
\bea
\nu_{j_o+1} &=&  \left(\frac{M}{2}-2kN_e-2kj_o\right ) \times P_x(F|E)]^{2k-1} \nonumber \\
& \times & P_y(G|E) \times [P(H|G \cap F)]^{2k-1}.
\eea
Substituting for each of the quantities on the right hand side, we obtain
the total number of ways of placing the $N_o$ rectangles on the odd
rows as 
\bea
\Omega_o &=& \frac{1}{N_o!}\prod_{j_o=0}^{N_o-1} \nu_{j_o+1} \nonumber \\
&=& \frac{1}{N_o!}\prod_{j_o=0}^{N_o-1} \frac{(\frac{M}{2}-2kN_e-2kj_o)^{4k}}
 {(\frac{M}{2}-2kj_o)^{2k}} \nonumber \\
&& \times \frac{(\frac{M}{2}-2kj_o+j_o)^{2k-1}}{(\frac{M}{2}-2kN_e-2kj_o+N_e+j_o)^{4k-2}}. 
\label{eq:omegaodd}
\eea

We would like to express the entropy in terms of the total density $\rho$ and 
the densities of occupied sites in even and odd rows,
given by $\rho_e$ and $\rho_o$ respectively.
Clearly,
\bea
\rho_e &= & \frac{4kN_e}{M},\\
\rho_o &= & \frac{4kN_o}{M}, \\
\rho&=& \rho_e+\rho_o.
\eea
The entropy  per site $s(\rho_e,\rho_o)$ in the thermodynamic limit is given by
\be
s(\rho_e, \rho_o) = \lim_{M\to\infty}\frac{1}{M} \ln \left(\Omega_0 \Omega_e \right).
\ee
Substituting for $\Omega_e$ and $\Omega_o$ from
Eqs.~(\ref{eq:omegaeven}) and (\ref{eq:omegaodd}), 
we obtain
\begin{widetext}
\bea 
s(\rho_e, \rho_o)& =& 
-\sum_{i=o,e}\frac{\rho_i}{4k}\ln \frac{\rho_i}{2k}-
(1- \rho) \ln(1- \rho) 
+\left (1-\rho+ \frac{\rho }{2k} \right) 
\ln \left(1-\rho+ \frac{\rho }{2k} \right) 
\nonumber \\ &&
+\sum_{i=o,e} \frac{1-\rho_i}{2} \ln (1-\rho_i)
-\frac{1}{2}  \sum_{i=o,e} \left ( 1-\rho_i+\frac{\rho_i}{2k} \right) 
\ln\left ( 1-\rho_i+\frac{\rho_i}{2k} \right).
\eea
\end{widetext}

\begin{figure}
\includegraphics[width=\columnwidth]{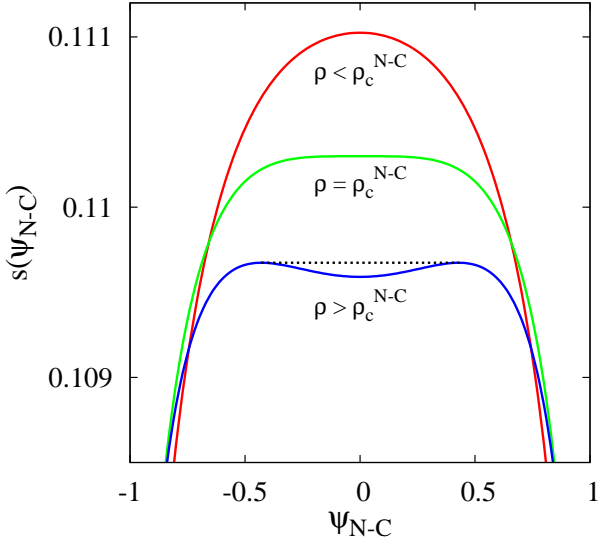}
\caption{(Color online) Entropy as a function as a function of the 
order parameter $\psi_{N-C}$ for m=2 near the N-C transition ($\rho_c^{N-C} \approx$ 0.624 ). 
The data are for $k=4$. The dotted line denotes the concave envelope.}
\label{fig:s_q_nc_m2}
\end{figure}

We express the entropy $s(\rho_e, \rho_o)$ in terms of density  $\rho$ and the 
order parameter $\psi_{N-C}$, defined as
\be
\psi_{N-C}=\frac{\rho_e-\rho_o}{\rho}.
\ee 
$\psi_{N-C}$ is zero in the nematic phase and non-zero in the columnar phase.
For a fixed value of $\rho$, the equilibrium values of $\rho_o$ and $\rho_e$ are
determined by maximizing the entropy
$s(\rho, \psi_{N-C})$ with respect to $\psi_{N-C}$. In Fig.~\ref{fig:s_q_nc_m2} we 
show the variation of entropy $s(\rho,\psi_{N-C})$ with $\psi_{N-C}$
for different densities. 
For small values of $\rho$, the
entropy is maximized by $\psi_{N-C}=0$, $i.e$, $\rho_e=\rho_o$. Beyond a critical density
$\rho_c^{N-C}$, $s(\rho,\psi_{N-C})$ is maximized by $\psi_{N-C}\neq 0$, $i.e$,  
$\rho_e \neq \rho_o$. $\psi_{N-C}$ grows continuously with $\rho$ for $\rho> \rho_c$,
and thus the transition for $m=2$ is continuous.

The expansion of $s(\rho, \psi_{N-C})$ in powers of  $\psi_{N-C}$ has only even powers
of $\psi_{N-C}$ since $s(\rho, \psi_{N-C})$ is invariant when 
$\psi_{N-C} \leftrightarrow
-\psi_{N-C}$. Thus, the critical density
is obtained from the condition $d^2 s /d \psi_{N-C}^2|_{\psi_{N-C}=0}=0$.
This gives
\bea
\rho_c^{N-C} &=& \frac{-1+4 k-\sqrt{1-4 k+8 k^2}}{2k-1} \nonumber \\
&=& (2-\sqrt{2})+\frac{A}{k}+O(k^{-2}),
\label{eq:rhocm2}
\eea
where $A=(\frac{1}{2}-\frac{1}{2\sqrt 2}) > 0$. We note that as $k\rightarrow \infty$, $\rho_c^{N-C}$ tends to a $k$
independent value and that the  transition exists for all $k \geq 2$. 

We can similarly calculate the entropy for $m=3$ and then generalize the 
expression of entropy for arbitrary $m$ and $k$.
Now, there are $m$ densities $\rho_1$, $\rho_2$, $\ldots$, $\rho_m$, 
corresponding to the $m$ types of rows. In terms of them,
the entropy is given by
\begin{widetext}
\bea 
s(\{\rho_i\})& =&
- \sum_{i=1}^{m} \frac{\rho_i}{m^2 k} \ln \frac{\rho_i}{m k}
- (1-\rho) \ln(1-  \rho)
+\left(1-\rho+\frac{\rho}{m k} \right) \ln \left(1-\rho+\frac{\rho}{m k}\right)
\nonumber \\ &&
+\sum_{i=1}^m \frac{1-\rho+\rho_i}{m}
\ln  \left(1-\rho+\rho_i \right) 
-\frac{1}{m} \sum_{i=1}^m \left(1-\rho+\rho_i + \frac{\rho-\rho_i}{m k}\right) 
\ln \left(1-\rho+\rho_i + \frac{\rho-\rho_i}{m k}\right).
\eea
\end{widetext}

\begin{figure}
\includegraphics[width=\columnwidth]{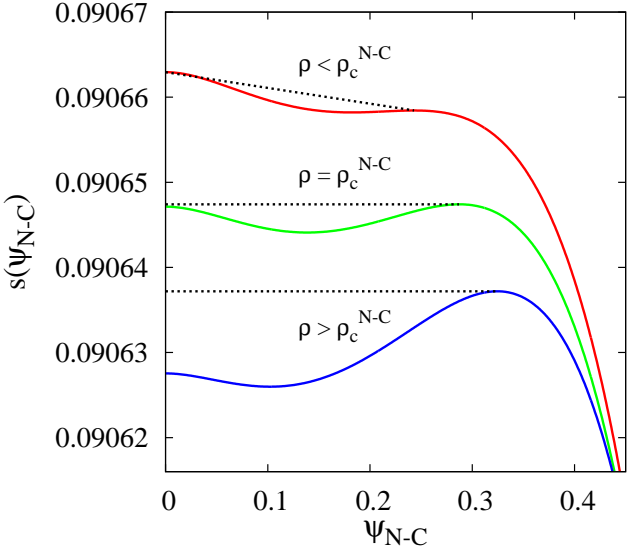}
\caption{(Color online) Entropy as a function as a function of the 
order parameter $\psi_{N-C}$ for m=3 near the N-C transition ($\rho_c^{N-C}\approx 0.684$). 
The data are for $k=2$. 
The dotted lines denote the concave envelopes.
The curves have been shifted for clarity. }
\label{fig:s_q_nc_m3}
\end{figure}

Here, we define the order parameter to be 
\be
\psi_{N-C}=\frac{\rho_1-\rho_2}{\rho},
\ee 
where we set $\rho_2=\rho_3=\ldots=\rho_m$. Now, $s(\psi_{N-C},\rho)$ is not invariant
when $\psi_{N-C}= -\psi_{N-C}$. Thus, when expanded in powers of $\psi_{N-C}$, $s(\psi_{N-C},\rho)$
has cubic terms, making the transition first order. This is illustrated in 
Fig.~\ref{fig:s_q_nc_m3} which shows the variation of entropy with $\psi_{N-C}$ for different
$\rho$ near the N-C transition.
For low densities $s(\psi_{N-C})$ exhibit a single peak at $\psi_{N-C}=0$, but with 
increasing $\rho$ a secondary maximum gets developed at $
\psi_{N-C} \neq 0$. For $\rho=\rho_c^{N-C}$ the 
maximum at $\psi_{N-C}=0$ and $\psi_{N-C}\neq 0$ becomes of equal height. Beyond 
$\rho_c^{N-C}$ the global maximum of $s(\psi_{N-C},\rho)$ jumps to $\psi_{N-C}\neq 0$, 
making the N-C transition to be first order.

\begin{figure}
\includegraphics[width=\columnwidth]{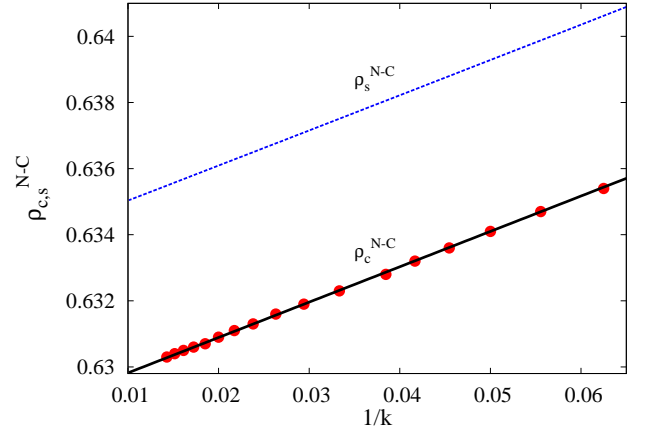}
\caption{(Color online)  The critical density $\rho_c^{N-C}$ and the spinodal density $\rho_s^{N-C}$, obtained from the Bethe approximation,
as a function of $1/k$ for $m=3$. 
The solid line is $0.62875+0.107/k$.}
\label{fig:rhoc_k_m3.eps}
\end{figure}

Unlike the $m=2$ case, there is no way to obtain an analytic
expression for $\rho_c^{N-C}$. For m=3, the numerically determined $\rho_c^{N-C}$ 
for different $k$  is shown in Fig.~(\ref{fig:rhoc_k_m3.eps}). From the data, we obtain
\be
\rho_c^{N-C} = 0.62875+\frac{0.107}{k}, \quad m=3.
\ee
We note that this expression has the same form as for $m=2$ [see Eq.~(\ref{eq:rhocm2})].

For $m>3$, we proceed as follows. The transition density $\rho_c^{N-C}$
is bounded from above by the
spinodal density $\rho_s^{N-C}$, the density at which the entropy at 
$\psi_{N-C}=0$ changes
from a local maximum to local minimum. $\rho_s^{N-C}$ is obtained from the condition
$d^2 s /d \psi_{N-C}^2|_{\psi_{N-C}=0}=0$ and we obtain
\bea
\rho^{N-C}_s &=& \frac{-m+ 2 k m^2 - m\sqrt{1-4 k + 4 k^2 m}}{2(1-m-k
m+k m^2)} \geq \rho_c^{N-C}, \\
&=&
\frac{\sqrt{m}}{1+\sqrt{m}}+\frac{B_1(m)}{k}+O\left(k^{-2}\right),~k
\to \infty,\\
&=& 1-\frac{1}{\sqrt{m}}+\frac{B_2(k)}{m}+O \left( m^{-3/2}\right),~ m
\to \infty,
\label{eq:spinodal}
\eea
where $B_1(m)=\frac{1}{2(m+\sqrt{m})}$ and $B_2(k)=(1+\frac{1}{2k})$. 
The spinodal density is compared with $\rho_c^{N-C}$ in 
Fig.~\ref{fig:rhoc_k_m3.eps}. 
From Eq.~(\ref{eq:spinodal}), it follows that 
$\rho_{s}^{N-C} \leq 1$ and tends to one when $m,k \to \infty$. The limit $m \to \infty$
corresponds to the continuum limit. In this limit $\rho_s^{N-C} \to 1$.
Thus, it is not clear whether the nematic-columnar phase transition 
will exist in the continuum.

\subsection{The columnar--sublattice phase boundary}

The dependence of the C-S phase boundary on $m$ and $k$ may be
determined by estimating the entropy for the C and S phases close to full packing. 
We approximate the
entropy of the C phase by the entropy of the fully aligned C phase. Since the heads
of the rectangles are all in either even or odd rows/columns, by ignoring the unoccupied
rows/columns,
the calculation of entropy reduces to an one dimensional problem of rods. 
The mean number of holes in a row is $L(1-\rho)$, and the mean number of 
rods (of length $m k$)   in a row is  $\frac{\rho L}{mk}$. There are $L/m$ such rows.
The number of ways of arranging the rods and holes on a row is
\be
\Omega_{row}=\frac{[L(1-\rho)+\frac{\rho L}{mk}]!}{[L(1-\rho)]![\frac{\rho L}{mk}]!},
\ee
such that the total number of ways of arranging the rectangles is $\Omega_{row}^{L/m}$.
Hence, ${\mathcal S}_C$,  the entropy per site of the columnar phase is given by
${\mathcal S}_C=(L m)^{-1} \ln \Omega_{row}$, which for 
densities close to $1$ is 
\be
\mathcal{S}_{C}\approx \frac{1-\rho}{m} \ln \left[\frac{e}{k m (1-\rho)}\right]+O[(1-\rho)^2].
\label{eq:s_smec}
\ee

We now estimate the entropy for the sublattice phase.
At full packing, the head of each rectangle is on one of $m^2$ sublattices.
Ignoring the sites belonging to other sublattices, it is easy to see that
each configuration of rectangles
can be mapped on to a configuration of rods
of length $k$ on a lattice of size $L/m \times L/m$. When 
$k \gg 1$, by solving the full packed problem of rods on strips, it is known 
that the entropy per unit site is 
$k^{-2} \ln k$~\cite{ghosh2007}.
For rectangles of size $m\times mk$ we obtain
\be 
\mathcal{S}_{S} (\rho=1) \approx \frac{\ln k}{m^2 k^2}, ~k\gg 1,
\ee
where $\mathcal{S}_S$ is the entropy per site of the S phase. For densities close to 1 
($\rho=1-\epsilon$) we  estimate the correction term to  $\mathcal{S}_{S}(\rho=1)$ by 
removing $\epsilon/(m^2 k)$ fraction of rectangles at random from the fully packed state. Here,
we ignore the entropy of the holes, assuming  that the holes form bound states. This
gives the  entropy of 
the sublattice phase, close to the full packing to be approximately
\be 
\mathcal{S}_{S}\approx \frac{\ln k}{m^2 k^2} -\frac{1}{m^2 k}[(1-\rho) \ln (1-\rho) + 
\rho \ln \rho], ~~k \gg 1,
\label{eq:s_sub}
\ee
Comparing Eqs.~(\ref{eq:s_smec}) and (\ref{eq:s_sub}) up to the leading 
order we obtain
the critical density for the C-S transition to be 
\be
\rho_c^{C-S} \approx 1-\frac{A_2}{m k^2}, ~k \gg 1,
\ee 
where $A_2>0$ is a constant.

Given the above asymptotic behavior of the phase boundaries, we expect that the phase
diagram for $m\geq 4$ will be similar to that obtained for $m=3$ for large $k$. For
small $k$, we check that for rectangles of size
$4\times 8$, there are two transitions just as in $3 \times 6$ rectangles. Thus,
we are led to conjecture that the phase diagram for $m\geq 4$ will be qualitatively similar
to that for $m=3$ for all $k$.

\section{\label{sec:summary}Summary and Discussion }

To summarize, we obtained the rich phase diagram of a system of $m 
\times m k$ hard rectangles on a square lattice for integer $m, k$ using 
a combination of Monte Carlo simulations and analytical calculations. 
We improve an existing cluster Monte Carlo algorithm by implementing 
the plaquette flip move, which reduces the autocorrelation time considerably. 
For $k\geq 7$, we show that the system undergoes three entropy driven 
transitions with increasing density. For $m=2$, we find that the I-N, 
N-C and C-S transitions are continuous, but the I-C transition may be 
continuous or first order, depending on $k$. The critical exponents for 
the continuous transitions were obtained using finite size scaling. The 
I-N and N-C transitions are found to be in the Ising universality class, 
while the C-S transition in the Ashkin-Teller universality class. The 
I-C transitions are also found to be in the Ashkin-Teller universality 
class when continuous. For larger $m$, the number of possible ordered 
states increase and the corresponding transitions become first order.

Surprisingly, our numerical data suggests that the nematic--columnar 
phase transition for $m=3$ is first order. However, once a nematic phase 
with orientational order exists, there are only three possible choices 
for the columnar phase. By analogy with the three state Potts model, we 
would expect a continuous transition, in contradiction with the 
numerical result. We also performed simulations for a system where the 
activity for vertical rectangles is zero (only horizontal rectangles are 
present) and observed again a first order transition. However, for 
$3\times 21$ rectangles, the autocorrelation time is high and it becomes 
increasingly difficult to obtain reliable data. Simulations of larger 
systems are required to resolve this puzzle in the future.

When $m=2$, the I-C transition is found to be continuous for $k=4$ and 
$5$, but first order for $k=6$. For $k=6$, it is a weak first order 
transition and it is difficult to see the jump in density for small 
system sizes. It is also possible that the data is
difficult to interpret because the transition point is  close to a
triciritcal point (the intersection of the I-N and N-C phase
boundaries). It would be interesting to reconfirm the first order 
nature by either simulating larger systems or doing constant density 
Monte Carlo simulations at the transition point so that phase separation 
may be seen. Also, determining a method to map $k$ and $\rho$ to the 
Ashkin Teller model parameters would be useful in clarifying this issue.

Another issue that we are not able to resolve completely is the 
determination of the minimum value of $k$ (say $k_{min}$) for which two 
transitions exist. For $m \geq 3$, we show that $k_{min}=2$. When $m=2$, 
our numerical data suggests that $k_{min}=4$, with a direct transition 
from isotropic to sublattice phase for $k=2,3$. However, for $k=4$, the 
columnar phase exists in a very narrow window of $\mu$ or $\rho$. 
Whether the columnar phase is present for $k=2, 3$, but we are unable to 
resolve the transitions, is something that requires investigation of 
much larger system sizes.

We obtained the N-C phase boundary analytically through a Bethe 
approximation. However, an improvement of the calculations is desirable 
as the approximations are ad hoc and uncontrolled and there does not 
appear to be a systematic way of improving it. An exact solution on tree 
like lattices, for example the random locally layered tree like lattice 
(RLTL)~\cite{dhar2011,joyjit_rltl2013} would be more satisfying. At 
present, we have not been able to formulate the problem of rectangles on 
RLTL. This is a promising area for future study.

Several extensions of the problem are possible. An interesting limit is 
the continuum problem of oriented rectangles of aspect ration $k$. This 
corresponds to the $m\to \infty$ limit of the lattice model. From the 
analytical arguments presented in the paper, we would expect a isotropic 
nematic transition at a critical density proportional to $k^{-1}$. 
Within the Bethe approximation, the spinodal density for the 
nematic--columnar transition tends to $1$ in the continuum limit. This 
being an upper bound for the critical density for the nematic--columnar 
transition, it is likely that the continuum problem will have a second 
transition into a columnar like phase, a true 
columnar phase being 
prohibited by the Mermin Wagner theorem. We expect the C-S transition to 
be absent in the continuum, as the critical density for the C-S 
transition tends to $1$ as $m\to \infty$. It would be interesting to 
verify these claims numerically. Preliminary simulations show an 
isotropic nematic transition.

One could also consider the model on other lattices like the triangular 
lattice. Here, the parallelograms may  orient themselves 
along three possible lattice directions. For each orientation 
the rectangles may have two different slants (as the shorter side of the 
rectangle may be oriented along two possible lattice directions). 
In this case, we expect the phase diagram to be qualitatively 
similar to that for the square lattice. However, as there are three 
broken symmetry phases corresponding to the three directions, the I-N 
transition would be in the 3 state Potts universality class, as opposed 
to Ising universality class for the square lattice. However, the N-C 
transition would be same as that for the square lattice. The I-C 
transition is now expected to be first order for all $m$ since the 
number of symmetric C phases increases from $2 m$ to $3 m$. 
The S phase will now have $2 m^2$ symmetric states, the extra 
factor of $2$ being due to the two different slants corresponding to each 
orientation. 
Therefore, we expect the I-S transition to 
be always first order. 
It would be 
interesting to study the C-S transition carefully in detail for the triangular lattice 
and compare with that for the square lattice.

We have also studied systems with non-integer $k$ on the square lattice 
(e.g., $2 \times 11$). Now, it is straightforward to see that the high 
density phase cannot have sublattice order. Thus, the high density phase 
does not possess either translational or orientational order. But, at 
intermediate densities, we observe the existence of the C phase for $k 
\geq k_{min}$ and N phase for $k\geq 7$. When $m=2$ we observe the C 
phase for $k\geq 5.5$ (rectangles of size $2\times 11$). In this case 
the I-C transition is found to be first order. For both $m=2$ and $3$, 
the I-N and N-C transitions are expected to be similar to that for 
integer $k$. Unlike integer $k$, we do not observe any transition for 
small $k$ ($k<5.5$ for $m=2$) and the phase remains isotropic at all 
densities. We hope to clarify these issues in detail in a future paper.

We argued, based on an analogy with the continuum problem, that 
$\rho_c^{I-N} \approx A_1/k$, where $A_1$ is independent of $m$. Can this 
conjecture be verified analytically or through numerical simulations? We 
expect that for $m=2$, the critical density can be determined 
numerically up to a $k$  large enough 
to determine $A_1$. Comparison with 
the value of $A_1$ for $m=1$~\cite{ghosh2007,fernandez2008c} would help in 
verifying the conjecture.

Extension to three dimensional cubic lattice would result in a much 
richer phase diagram that remains to be explored. The algorithm used in 
this paper is easily implementable in three dimensions.

Finally, the $m=1$ case (hard rods) is the only instance where the 
existence of a nematic phase may be proved 
rigorously~\cite{giuliani2013}. To the best of our knowledge, there 
exists no proof of existence of phases with partial translational order 
like the columnar phase. The hard rectangle model seems an ideal 
candidate to prove its existence.

\begin{acknowledgments}
We thank Deepak Dhar and J\"{u}rgen F. Stilck for helpful
discussions.
The simulations were carried out on the supercomputing
machine Annapurna at The Institute of Mathematical Sciences. 
\end{acknowledgments}


\begin{thebibliography}{55}%
\makeatletter
\providecommand \@ifxundefined [1]{%
 \@ifx{#1\undefined}
}%
\providecommand \@ifnum [1]{%
 \ifnum #1\expandafter \@firstoftwo
 \else \expandafter \@secondoftwo
 \fi
}%
\providecommand \@ifx [1]{%
 \ifx #1\expandafter \@firstoftwo
 \else \expandafter \@secondoftwo
 \fi
}%
\providecommand \natexlab [1]{#1}%
\providecommand \enquote  [1]{``#1''}%
\providecommand \bibnamefont  [1]{#1}%
\providecommand \bibfnamefont [1]{#1}%
\providecommand \citenamefont [1]{#1}%
\providecommand \href@noop [0]{\@secondoftwo}%
\providecommand \href [0]{\begingroup \@sanitize@url \@href}%
\providecommand \@href[1]{\@@startlink{#1}\@@href}%
\providecommand \@@href[1]{\endgroup#1\@@endlink}%
\providecommand \@sanitize@url [0]{\catcode `\\12\catcode `\$12\catcode
  `\&12\catcode `\#12\catcode `\^12\catcode `\_12\catcode `\%12\relax}%
\providecommand \@@startlink[1]{}%
\providecommand \@@endlink[0]{}%
\providecommand \url  [0]{\begingroup\@sanitize@url \@url }%
\providecommand \@url [1]{\endgroup\@href {#1}{\urlprefix }}%
\providecommand \urlprefix  [0]{URL }%
\providecommand \Eprint [0]{\href }%
\providecommand \doibase [0]{http://dx.doi.org/}%
\providecommand \selectlanguage [0]{\@gobble}%
\providecommand \bibinfo  [0]{\@secondoftwo}%
\providecommand \bibfield  [0]{\@secondoftwo}%
\providecommand \translation [1]{[#1]}%
\providecommand \BibitemOpen [0]{}%
\providecommand \bibitemStop [0]{}%
\providecommand \bibitemNoStop [0]{.\EOS\space}%
\providecommand \EOS [0]{\spacefactor3000\relax}%
\providecommand \BibitemShut  [1]{\csname bibitem#1\endcsname}%
\let\auto@bib@innerbib\@empty
\bibitem [{\citenamefont {Onsager}(1949)}]{onsager1949}%
  \BibitemOpen
  \bibfield  {author} {\bibinfo {author} {\bibfnamefont {L.}~\bibnamefont
  {Onsager}},\ }\href@noop {} {\bibfield  {journal} {\bibinfo  {journal} {Ann.
  N.Y. Acad. Sci.}\ }\textbf {\bibinfo {volume} {51}},\ \bibinfo {pages} {627}
  (\bibinfo {year} {1949})}\BibitemShut {NoStop}%
\bibitem [{\citenamefont {Flory}(1956)}]{flory1956b}%
  \BibitemOpen
  \bibfield  {author} {\bibinfo {author} {\bibfnamefont {P.~J.}\ \bibnamefont
  {Flory}},\ }\href@noop {} {\bibfield  {journal} {\bibinfo  {journal} {Proc.
  R. Soc.}\ }\textbf {\bibinfo {volume} {234}},\ \bibinfo {pages} {73}
  (\bibinfo {year} {1956})}\BibitemShut {NoStop}%
\bibitem [{\citenamefont {de~Gennes}\ and\ \citenamefont
  {Prost}(1995)}]{degennesBook}%
  \BibitemOpen
  \bibfield  {author} {\bibinfo {author} {\bibfnamefont {P.~G.}\ \bibnamefont
  {de~Gennes}}\ and\ \bibinfo {author} {\bibfnamefont {J.}~\bibnamefont
  {Prost}},\ }\href@noop {} {\emph {\bibinfo {title} {The Physics of Liquid
  Crystals}}}\ (\bibinfo  {publisher} {Oxford University Press},\ \bibinfo
  {address} {Oxford},\ \bibinfo {year} {1995})\ pp.\ \bibinfo {pages}
  {64--66}\BibitemShut {NoStop}%
\bibitem [{\citenamefont {Vroege}\ and\ \citenamefont
  {Lekkerkerker}(1992)}]{vroege1992}%
  \BibitemOpen
  \bibfield  {author} {\bibinfo {author} {\bibfnamefont {G.~J.}\ \bibnamefont
  {Vroege}}\ and\ \bibinfo {author} {\bibfnamefont {H.~N.~W.}\ \bibnamefont
  {Lekkerkerker}},\ }\href@noop {} {\bibfield  {journal} {\bibinfo  {journal}
  {Rep. Prog. Phys.}\ }\textbf {\bibinfo {volume} {55}},\ \bibinfo {pages}
  {1241} (\bibinfo {year} {1992})}\BibitemShut {NoStop}%
\bibitem [{\citenamefont {Dhar}\ \emph {et~al.}(2011)\citenamefont {Dhar},
  \citenamefont {Rajesh},\ and\ \citenamefont {Stilck}}]{dhar2011}%
  \BibitemOpen
  \bibfield  {author} {\bibinfo {author} {\bibfnamefont {D.}~\bibnamefont
  {Dhar}}, \bibinfo {author} {\bibfnamefont {R.}~\bibnamefont {Rajesh}}, \ and\
  \bibinfo {author} {\bibfnamefont {J.~F.}\ \bibnamefont {Stilck}},\
  }\href@noop {} {\bibfield  {journal} {\bibinfo  {journal} {Phys. Rev. E}\
  }\textbf {\bibinfo {volume} {84}},\ \bibinfo {pages} {011140} (\bibinfo
  {year} {2011})}\BibitemShut {NoStop}%
\bibitem [{\citenamefont {Frenkel}\ \emph {et~al.}(1988)\citenamefont
  {Frenkel}, \citenamefont {Lekkerkerker},\ and\ \citenamefont
  {Stroobants}}]{frenkel1988}%
  \BibitemOpen
  \bibfield  {author} {\bibinfo {author} {\bibfnamefont {D.}~\bibnamefont
  {Frenkel}}, \bibinfo {author} {\bibfnamefont {H.~N.~W.}\ \bibnamefont
  {Lekkerkerker}}, \ and\ \bibinfo {author} {\bibfnamefont {A.}~\bibnamefont
  {Stroobants}},\ }\href@noop {} {\bibfield  {journal} {\bibinfo  {journal}
  {Nature}\ }\textbf {\bibinfo {volume} {332}},\ \bibinfo {pages} {822}
  (\bibinfo {year} {1988})}\BibitemShut {NoStop}%
\bibitem [{\citenamefont {Bolhuis}\ and\ \citenamefont
  {Frenkel}(1997)}]{frenkel1997}%
  \BibitemOpen
  \bibfield  {author} {\bibinfo {author} {\bibfnamefont {P.}~\bibnamefont
  {Bolhuis}}\ and\ \bibinfo {author} {\bibfnamefont {D.}~\bibnamefont
  {Frenkel}},\ }\href@noop {} {\bibfield  {journal} {\bibinfo  {journal} {J.
  Chem. Phys}\ }\textbf {\bibinfo {volume} {106}},\ \bibinfo {pages} {666}
  (\bibinfo {year} {1997})}\BibitemShut {NoStop}%
\bibitem [{\citenamefont {Straley}(1971)}]{straley1971}%
  \BibitemOpen
  \bibfield  {author} {\bibinfo {author} {\bibfnamefont {J.~P.}\ \bibnamefont
  {Straley}},\ }\href@noop {} {\bibfield  {journal} {\bibinfo  {journal} {Phys.
  Rev. A}\ }\textbf {\bibinfo {volume} {4}},\ \bibinfo {pages} {675} (\bibinfo
  {year} {1971})}\BibitemShut {NoStop}%
\bibitem [{\citenamefont {Frenkel}\ and\ \citenamefont
  {Eppenga}(1985)}]{frenkel1985}%
  \BibitemOpen
  \bibfield  {author} {\bibinfo {author} {\bibfnamefont {D.}~\bibnamefont
  {Frenkel}}\ and\ \bibinfo {author} {\bibfnamefont {R.}~\bibnamefont
  {Eppenga}},\ }\href@noop {} {\bibfield  {journal} {\bibinfo  {journal} {Phys.
  Rev. A}\ }\textbf {\bibinfo {volume} {31}},\ \bibinfo {pages} {1776}
  (\bibinfo {year} {1985})}\BibitemShut {NoStop}%
\bibitem [{\citenamefont {Wojciechowski}\ and\ \citenamefont
  {Frenkel}(2004)}]{frenkel2004}%
  \BibitemOpen
  \bibfield  {author} {\bibinfo {author} {\bibfnamefont {K.~W.}\ \bibnamefont
  {Wojciechowski}}\ and\ \bibinfo {author} {\bibfnamefont {D.}~\bibnamefont
  {Frenkel}},\ }\href@noop {} {\bibfield  {journal} {\bibinfo  {journal}
  {Comput. Methods Sci. Tech.}\ }\textbf {\bibinfo {volume} {10}},\ \bibinfo
  {pages} {235} (\bibinfo {year} {2004})}\BibitemShut {NoStop}%
\bibitem [{\citenamefont {Khandkar}\ and\ \citenamefont
  {Barma}(2005)}]{khandkar2005}%
  \BibitemOpen
  \bibfield  {author} {\bibinfo {author} {\bibfnamefont {M.~D.}\ \bibnamefont
  {Khandkar}}\ and\ \bibinfo {author} {\bibfnamefont {M.}~\bibnamefont
  {Barma}},\ }\href@noop {} {\bibfield  {journal} {\bibinfo  {journal} {Phys.
  Rev. E}\ }\textbf {\bibinfo {volume} {72}},\ \bibinfo {pages} {051717}
  (\bibinfo {year} {2005})}\BibitemShut {NoStop}%
\bibitem [{\citenamefont {Donev}\ \emph {et~al.}(2006)\citenamefont {Donev},
  \citenamefont {Burton}, \citenamefont {Stillinger},\ and\ \citenamefont
  {Torquato}}]{donev2006}%
  \BibitemOpen
  \bibfield  {author} {\bibinfo {author} {\bibfnamefont {A.}~\bibnamefont
  {Donev}}, \bibinfo {author} {\bibfnamefont {J.}~\bibnamefont {Burton}},
  \bibinfo {author} {\bibfnamefont {F.~H.}\ \bibnamefont {Stillinger}}, \ and\
  \bibinfo {author} {\bibfnamefont {S.}~\bibnamefont {Torquato}},\ }\href@noop
  {} {\bibfield  {journal} {\bibinfo  {journal} {Phys. Rev. B}\ }\textbf
  {\bibinfo {volume} {73}},\ \bibinfo {pages} {054109} (\bibinfo {year}
  {2006})}\BibitemShut {NoStop}%
\bibitem [{\citenamefont {Zhao}\ \emph {et~al.}(2007)\citenamefont {Zhao},
  \citenamefont {Harrison}, \citenamefont {Huse}, \citenamefont {Russel},\ and\
  \citenamefont {Chaikin}}]{zhao2007}%
  \BibitemOpen
  \bibfield  {author} {\bibinfo {author} {\bibfnamefont {K.}~\bibnamefont
  {Zhao}}, \bibinfo {author} {\bibfnamefont {C.}~\bibnamefont {Harrison}},
  \bibinfo {author} {\bibfnamefont {D.}~\bibnamefont {Huse}}, \bibinfo {author}
  {\bibfnamefont {W.~B.}\ \bibnamefont {Russel}}, \ and\ \bibinfo {author}
  {\bibfnamefont {P.~M.}\ \bibnamefont {Chaikin}},\ }\href@noop {} {\bibfield
  {journal} {\bibinfo  {journal} {Phys. Rev. E}\ }\textbf {\bibinfo {volume}
  {76}},\ \bibinfo {pages} {040401} (\bibinfo {year} {2007})}\BibitemShut
  {NoStop}%
\bibitem [{\citenamefont {Vink}(2009)}]{vink2009}%
  \BibitemOpen
  \bibfield  {author} {\bibinfo {author} {\bibfnamefont {R.~L.~C.}\
  \bibnamefont {Vink}},\ }\href@noop {} {\bibfield  {journal} {\bibinfo
  {journal} {Euro. Phys. J. B}\ }\textbf {\bibinfo {volume} {72}},\ \bibinfo
  {pages} {225} (\bibinfo {year} {2009})}\BibitemShut {NoStop}%
\bibitem [{\citenamefont {Fraden}\ \emph {et~al.}(1989)\citenamefont {Fraden},
  \citenamefont {Maret}, \citenamefont {Caspar},\ and\ \citenamefont
  {Meyer}}]{fraden1989}%
  \BibitemOpen
  \bibfield  {author} {\bibinfo {author} {\bibfnamefont {S.}~\bibnamefont
  {Fraden}}, \bibinfo {author} {\bibfnamefont {G.}~\bibnamefont {Maret}},
  \bibinfo {author} {\bibfnamefont {D.~L.~D.}\ \bibnamefont {Caspar}}, \ and\
  \bibinfo {author} {\bibfnamefont {R.~B.}\ \bibnamefont {Meyer}},\ }\href@noop
  {} {\bibfield  {journal} {\bibinfo  {journal} {Phys. Rev. Lett.}\ }\textbf
  {\bibinfo {volume} {63}},\ \bibinfo {pages} {2068} (\bibinfo {year}
  {1989})}\BibitemShut {NoStop}%
\bibitem [{\citenamefont {Islam}\ \emph {et~al.}(2004)\citenamefont {Islam},
  \citenamefont {Alsayed}, \citenamefont {Dogic}, \citenamefont {Zhang},
  \citenamefont {Lubensky},\ and\ \citenamefont {Yodh}}]{islam2004}%
  \BibitemOpen
  \bibfield  {author} {\bibinfo {author} {\bibfnamefont {M.~F.}\ \bibnamefont
  {Islam}}, \bibinfo {author} {\bibfnamefont {A.~M.}\ \bibnamefont {Alsayed}},
  \bibinfo {author} {\bibfnamefont {Z.}~\bibnamefont {Dogic}}, \bibinfo
  {author} {\bibfnamefont {J.}~\bibnamefont {Zhang}}, \bibinfo {author}
  {\bibfnamefont {T.~C.}\ \bibnamefont {Lubensky}}, \ and\ \bibinfo {author}
  {\bibfnamefont {A.~G.}\ \bibnamefont {Yodh}},\ }\href@noop {} {\bibfield
  {journal} {\bibinfo  {journal} {Phys. Rev. Lett.}\ }\textbf {\bibinfo
  {volume} {92}},\ \bibinfo {pages} {088303} (\bibinfo {year}
  {2004})}\BibitemShut {NoStop}%
\bibitem [{\citenamefont {Zhao}\ \emph {et~al.}(2011)\citenamefont {Zhao},
  \citenamefont {Bruinsma},\ and\ \citenamefont {Mason}}]{zhao2011}%
  \BibitemOpen
  \bibfield  {author} {\bibinfo {author} {\bibfnamefont {K.}~\bibnamefont
  {Zhao}}, \bibinfo {author} {\bibfnamefont {R.}~\bibnamefont {Bruinsma}}, \
  and\ \bibinfo {author} {\bibfnamefont {T.~G.}\ \bibnamefont {Mason}},\
  }\href@noop {} {\bibfield  {journal} {\bibinfo  {journal} {Proc. Natl. Acad.
  Sci.}\ }\textbf {\bibinfo {volume} {108}},\ \bibinfo {pages} {2684} (\bibinfo
  {year} {2011})}\BibitemShut {NoStop}%
\bibitem [{\citenamefont {Ghosh}\ and\ \citenamefont {Dhar}(2007)}]{ghosh2007}%
  \BibitemOpen
  \bibfield  {author} {\bibinfo {author} {\bibfnamefont {A.}~\bibnamefont
  {Ghosh}}\ and\ \bibinfo {author} {\bibfnamefont {D.}~\bibnamefont {Dhar}},\
  }\href@noop {} {\bibfield  {journal} {\bibinfo  {journal} {Euro. Phys.
  Lett.}\ }\textbf {\bibinfo {volume} {78}},\ \bibinfo {pages} {20003}
  (\bibinfo {year} {2007})}\BibitemShut {NoStop}%
\bibitem [{\citenamefont {Disertori}\ and\ \citenamefont
  {Giuliani}(2013)}]{giuliani2013}%
  \BibitemOpen
  \bibfield  {author} {\bibinfo {author} {\bibfnamefont {M.}~\bibnamefont
  {Disertori}}\ and\ \bibinfo {author} {\bibfnamefont {A.}~\bibnamefont
  {Giuliani}},\ }\href@noop {} {\bibfield  {journal} {\bibinfo  {journal}
  {Commun. Math. Phys.}\ }\textbf {\bibinfo {volume} {323}},\ \bibinfo {pages}
  {143} (\bibinfo {year} {2013})}\BibitemShut {NoStop}%
\bibitem [{\citenamefont {Kundu}\ \emph {et~al.}(2013)\citenamefont {Kundu},
  \citenamefont {Rajesh}, \citenamefont {Dhar},\ and\ \citenamefont
  {Stilck}}]{joyjit2013}%
  \BibitemOpen
  \bibfield  {author} {\bibinfo {author} {\bibfnamefont {J.}~\bibnamefont
  {Kundu}}, \bibinfo {author} {\bibfnamefont {R.}~\bibnamefont {Rajesh}},
  \bibinfo {author} {\bibfnamefont {D.}~\bibnamefont {Dhar}}, \ and\ \bibinfo
  {author} {\bibfnamefont {J.~F.}\ \bibnamefont {Stilck}},\ }\href@noop {}
  {\bibfield  {journal} {\bibinfo  {journal} {Phys. Rev. E}\ }\textbf {\bibinfo
  {volume} {87}},\ \bibinfo {pages} {032103} (\bibinfo {year}
  {2013})}\BibitemShut {NoStop}%
\bibitem [{\citenamefont {Matoz-Fernandez}\ \emph
  {et~al.}(2008{\natexlab{a}})\citenamefont {Matoz-Fernandez}, \citenamefont
  {Linares},\ and\ \citenamefont {Ramirez-Pastor}}]{fernandez2008a}%
  \BibitemOpen
  \bibfield  {author} {\bibinfo {author} {\bibfnamefont {D.~A.}\ \bibnamefont
  {Matoz-Fernandez}}, \bibinfo {author} {\bibfnamefont {D.~H.}\ \bibnamefont
  {Linares}}, \ and\ \bibinfo {author} {\bibfnamefont {A.~J.}\ \bibnamefont
  {Ramirez-Pastor}},\ }\href@noop {} {\bibfield  {journal} {\bibinfo  {journal}
  {Euro. Phys. Lett}\ }\textbf {\bibinfo {volume} {82}},\ \bibinfo {pages}
  {50007} (\bibinfo {year} {2008}{\natexlab{a}})}\BibitemShut {NoStop}%
\bibitem [{\citenamefont {Fischer}\ and\ \citenamefont
  {Vink}(2009)}]{fischer2009}%
  \BibitemOpen
  \bibfield  {author} {\bibinfo {author} {\bibfnamefont {T.}~\bibnamefont
  {Fischer}}\ and\ \bibinfo {author} {\bibfnamefont {R.~L.~C.}\ \bibnamefont
  {Vink}},\ }\href@noop {} {\bibfield  {journal} {\bibinfo  {journal} {Euro.
  Phys. Lett.}\ }\textbf {\bibinfo {volume} {85}},\ \bibinfo {pages} {56003}
  (\bibinfo {year} {2009})}\BibitemShut {NoStop}%
\bibitem [{\citenamefont {Kundu}\ and\ \citenamefont
  {Rajesh}(2013)}]{joyjit_rltl2013}%
  \BibitemOpen
  \bibfield  {author} {\bibinfo {author} {\bibfnamefont {J.}~\bibnamefont
  {Kundu}}\ and\ \bibinfo {author} {\bibfnamefont {R.}~\bibnamefont {Rajesh}},\
  }\href@noop {} {\bibfield  {journal} {\bibinfo  {journal} {Phys. Rev. E}\
  }\textbf {\bibinfo {volume} {88}},\ \bibinfo {pages} {012134} (\bibinfo
  {year} {2013})}\BibitemShut {NoStop}%
\bibitem [{\citenamefont {Baxter}(1982)}]{baxterBook}%
  \BibitemOpen
  \bibfield  {author} {\bibinfo {author} {\bibfnamefont {R.~J.}\ \bibnamefont
  {Baxter}},\ }\href@noop {} {\emph {\bibinfo {title} {Exactly Solved Models in
  Statistical Mechanics}}}\ (\bibinfo  {publisher} {Academic Press},\ \bibinfo
  {address} {London},\ \bibinfo {year} {1982})\BibitemShut {NoStop}%
\bibitem [{\citenamefont {Fernandes}\ \emph {et~al.}(2007)\citenamefont
  {Fernandes}, \citenamefont {Arenzon},\ and\ \citenamefont
  {Levin}}]{fernandes2007}%
  \BibitemOpen
  \bibfield  {author} {\bibinfo {author} {\bibfnamefont {H.~C.~M.}\
  \bibnamefont {Fernandes}}, \bibinfo {author} {\bibfnamefont {J.~J.}\
  \bibnamefont {Arenzon}}, \ and\ \bibinfo {author} {\bibfnamefont
  {Y.}~\bibnamefont {Levin}},\ }\href@noop {} {\bibfield  {journal} {\bibinfo
  {journal} {J. Chem. Phys.}\ }\textbf {\bibinfo {volume} {126}},\ \bibinfo
  {pages} {114508} (\bibinfo {year} {2007})}\BibitemShut {NoStop}%
\bibitem [{\citenamefont {Zhitomirsky}\ and\ \citenamefont
  {Tsunetsugu}(2007)}]{hirokazu2007}%
  \BibitemOpen
  \bibfield  {author} {\bibinfo {author} {\bibfnamefont {M.~E.}\ \bibnamefont
  {Zhitomirsky}}\ and\ \bibinfo {author} {\bibfnamefont {H.}~\bibnamefont
  {Tsunetsugu}},\ }\href@noop {} {\bibfield  {journal} {\bibinfo  {journal}
  {Phys. Rev. B}\ }\textbf {\bibinfo {volume} {75}},\ \bibinfo {pages} {224416}
  (\bibinfo {year} {2007})}\BibitemShut {NoStop}%
\bibitem [{\citenamefont {Feng}\ \emph {et~al.}(2011)\citenamefont {Feng},
  \citenamefont {Bl\"{o}te},\ and\ \citenamefont {Nienhuis}}]{nienhuis2011}%
  \BibitemOpen
  \bibfield  {author} {\bibinfo {author} {\bibfnamefont {X.}~\bibnamefont
  {Feng}}, \bibinfo {author} {\bibfnamefont {H.~W.~J.}\ \bibnamefont
  {Bl\"{o}te}}, \ and\ \bibinfo {author} {\bibfnamefont {B.}~\bibnamefont
  {Nienhuis}},\ }\href@noop {} {\bibfield  {journal} {\bibinfo  {journal}
  {Phys. Rev. E}\ }\textbf {\bibinfo {volume} {83}},\ \bibinfo {pages} {061153}
  (\bibinfo {year} {2011})}\BibitemShut {NoStop}%
\bibitem [{\citenamefont {Ramola}(2012)}]{kabirthesis}%
  \BibitemOpen
  \bibfield  {author} {\bibinfo {author} {\bibfnamefont {K.}~\bibnamefont
  {Ramola}},\ }\emph {\bibinfo {title} {Onset of Order in Lattice Systems:
  Kitaev Model and Hard Squares}},\ \href@noop {} {Ph.D. thesis},\ \bibinfo
  {school} {Tata Institute of Fundamental Research, Mumbai} (\bibinfo {year}
  {2012})\BibitemShut {NoStop}%
\bibitem [{\citenamefont {Zwanzig}(1963)}]{zwanzig1963}%
  \BibitemOpen
  \bibfield  {author} {\bibinfo {author} {\bibfnamefont {R.}~\bibnamefont
  {Zwanzig}},\ }\href@noop {} {\bibfield  {journal} {\bibinfo  {journal} {J.
  Chem. Phys.}\ }\textbf {\bibinfo {volume} {39}},\ \bibinfo {pages} {1714}
  (\bibinfo {year} {1963})}\BibitemShut {NoStop}%
\bibitem [{\citenamefont {Heilmann}\ and\ \citenamefont
  {Lieb}(1970)}]{Heilmann1970}%
  \BibitemOpen
  \bibfield  {author} {\bibinfo {author} {\bibfnamefont {O.~J.}\ \bibnamefont
  {Heilmann}}\ and\ \bibinfo {author} {\bibfnamefont {E.~H.}\ \bibnamefont
  {Lieb}},\ }\href@noop {} {\bibfield  {journal} {\bibinfo  {journal} {Phys.
  Rev. Lett.}\ }\textbf {\bibinfo {volume} {24}},\ \bibinfo {pages} {1412}
  (\bibinfo {year} {1970})}\BibitemShut {NoStop}%
\bibitem [{\citenamefont {Kunz}(1970)}]{Kunz1970}%
  \BibitemOpen
  \bibfield  {author} {\bibinfo {author} {\bibfnamefont {H.}~\bibnamefont
  {Kunz}},\ }\href@noop {} {\bibfield  {journal} {\bibinfo  {journal} {Phys.
  Lett. A}\ }\textbf {\bibinfo {volume} {32}},\ \bibinfo {pages} {311}
  (\bibinfo {year} {1970})}\BibitemShut {NoStop}%
\bibitem [{\citenamefont {Gruber}\ and\ \citenamefont
  {Kunz}(1971)}]{Gruber1971}%
  \BibitemOpen
  \bibfield  {author} {\bibinfo {author} {\bibfnamefont {C.}~\bibnamefont
  {Gruber}}\ and\ \bibinfo {author} {\bibfnamefont {H.}~\bibnamefont {Kunz}},\
  }\href@noop {} {\bibfield  {journal} {\bibinfo  {journal} {Commun. Math.
  Phys.}\ }\textbf {\bibinfo {volume} {22}},\ \bibinfo {pages} {133} (\bibinfo
  {year} {1971})}\BibitemShut {NoStop}%
\bibitem [{\citenamefont {Heilmann}\ and\ \citenamefont
  {Lieb}(1972)}]{lieb1972}%
  \BibitemOpen
  \bibfield  {author} {\bibinfo {author} {\bibfnamefont {O.~J.}\ \bibnamefont
  {Heilmann}}\ and\ \bibinfo {author} {\bibfnamefont {E.}~\bibnamefont
  {Lieb}},\ }\href@noop {} {\bibfield  {journal} {\bibinfo  {journal} {Commun.
  Math. Phys.}\ }\textbf {\bibinfo {volume} {25}},\ \bibinfo {pages} {190}
  (\bibinfo {year} {1972})}\BibitemShut {NoStop}%
\bibitem [{\citenamefont {Barnes}\ \emph
  {et~al.}(2009{\natexlab{a}})\citenamefont {Barnes}, \citenamefont
  {Siderius},\ and\ \citenamefont {Gelb}}]{barnes2009}%
  \BibitemOpen
  \bibfield  {author} {\bibinfo {author} {\bibfnamefont {B.~C.}\ \bibnamefont
  {Barnes}}, \bibinfo {author} {\bibfnamefont {D.~W.}\ \bibnamefont
  {Siderius}}, \ and\ \bibinfo {author} {\bibfnamefont {L.~D.}\ \bibnamefont
  {Gelb}},\ }\href@noop {} {\bibfield  {journal} {\bibinfo  {journal}
  {Langmuir}\ }\textbf {\bibinfo {volume} {25}},\ \bibinfo {pages} {6702}
  (\bibinfo {year} {2009}{\natexlab{a}})}\BibitemShut {NoStop}%
\bibitem [{\citenamefont {Casey}\ and\ \citenamefont
  {Harrowell}(1995)}]{casey1995}%
  \BibitemOpen
  \bibfield  {author} {\bibinfo {author} {\bibfnamefont {A.}~\bibnamefont
  {Casey}}\ and\ \bibinfo {author} {\bibfnamefont {P.}~\bibnamefont
  {Harrowell}},\ }\href@noop {} {\bibfield  {journal} {\bibinfo  {journal} {J.
  Chem. Phys.}\ }\textbf {\bibinfo {volume} {103}},\ \bibinfo {pages} {6143}
  (\bibinfo {year} {1995})}\BibitemShut {NoStop}%
\bibitem [{\citenamefont {Patrykiejew}\ \emph {et~al.}(2000)\citenamefont
  {Patrykiejew}, \citenamefont {Sokolowski},\ and\ \citenamefont
  {Binder}}]{binder2000}%
  \BibitemOpen
  \bibfield  {author} {\bibinfo {author} {\bibfnamefont {A.}~\bibnamefont
  {Patrykiejew}}, \bibinfo {author} {\bibfnamefont {S.}~\bibnamefont
  {Sokolowski}}, \ and\ \bibinfo {author} {\bibfnamefont {K.}~\bibnamefont
  {Binder}},\ }\href@noop {} {\bibfield  {journal} {\bibinfo  {journal} {Surf.
  Sci. Rep.}\ }\textbf {\bibinfo {volume} {37}},\ \bibinfo {pages} {207}
  (\bibinfo {year} {2000})}\BibitemShut {NoStop}%
\bibitem [{\citenamefont {Taylor}\ \emph {et~al.}(1985)\citenamefont {Taylor},
  \citenamefont {Williams}, \citenamefont {Park}, \citenamefont {Bartelt},\
  and\ \citenamefont {Einstein}}]{taylor1985}%
  \BibitemOpen
  \bibfield  {author} {\bibinfo {author} {\bibfnamefont {D.~E.}\ \bibnamefont
  {Taylor}}, \bibinfo {author} {\bibfnamefont {E.~D.}\ \bibnamefont
  {Williams}}, \bibinfo {author} {\bibfnamefont {R.~L.}\ \bibnamefont {Park}},
  \bibinfo {author} {\bibfnamefont {N.~C.}\ \bibnamefont {Bartelt}}, \ and\
  \bibinfo {author} {\bibfnamefont {T.~L.}\ \bibnamefont {Einstein}},\
  }\href@noop {} {\bibfield  {journal} {\bibinfo  {journal} {Phys. Rev. B}\
  }\textbf {\bibinfo {volume} {32}},\ \bibinfo {pages} {4653} (\bibinfo {year}
  {1985})}\BibitemShut {NoStop}%
\bibitem [{\citenamefont {Bak}\ \emph {et~al.}(1985)\citenamefont {Bak},
  \citenamefont {Kleban}, \citenamefont {Unertl}, \citenamefont {Ochab},
  \citenamefont {Akinci}, \citenamefont {Bartelt},\ and\ \citenamefont
  {Einstein}}]{bak1985}%
  \BibitemOpen
  \bibfield  {author} {\bibinfo {author} {\bibfnamefont {P.}~\bibnamefont
  {Bak}}, \bibinfo {author} {\bibfnamefont {P.}~\bibnamefont {Kleban}},
  \bibinfo {author} {\bibfnamefont {W.~N.}\ \bibnamefont {Unertl}}, \bibinfo
  {author} {\bibfnamefont {J.}~\bibnamefont {Ochab}}, \bibinfo {author}
  {\bibfnamefont {G.}~\bibnamefont {Akinci}}, \bibinfo {author} {\bibfnamefont
  {N.~C.}\ \bibnamefont {Bartelt}}, \ and\ \bibinfo {author} {\bibfnamefont
  {T.~L.}\ \bibnamefont {Einstein}},\ }\href@noop {} {\bibfield  {journal}
  {\bibinfo  {journal} {Phys. Rev. Lett.}\ }\textbf {\bibinfo {volume} {54}},\
  \bibinfo {pages} {1539} (\bibinfo {year} {1985})}\BibitemShut {NoStop}%
\bibitem [{\citenamefont {Clarke}\ \emph {et~al.}(2000)\citenamefont {Clarke},
  \citenamefont {Cuesta}, \citenamefont {Sear}, \citenamefont {Sollich},\ and\
  \citenamefont {Speranza}}]{sear2000}%
  \BibitemOpen
  \bibfield  {author} {\bibinfo {author} {\bibfnamefont {N.}~\bibnamefont
  {Clarke}}, \bibinfo {author} {\bibfnamefont {J.~A.}\ \bibnamefont {Cuesta}},
  \bibinfo {author} {\bibfnamefont {R.}~\bibnamefont {Sear}}, \bibinfo {author}
  {\bibfnamefont {P.}~\bibnamefont {Sollich}}, \ and\ \bibinfo {author}
  {\bibfnamefont {A.}~\bibnamefont {Speranza}},\ }\href@noop {} {\bibfield
  {journal} {\bibinfo  {journal} {J. Chem. Phys}\ }\textbf {\bibinfo {volume}
  {113}},\ \bibinfo {pages} {5817} (\bibinfo {year} {2000})}\BibitemShut
  {NoStop}%
\bibitem [{\citenamefont {Mart\'{i}nez-Rat\'{o}n}\ and\ \citenamefont
  {Cuesta}(2003)}]{cuesta2003}%
  \BibitemOpen
  \bibfield  {author} {\bibinfo {author} {\bibfnamefont {Y.}~\bibnamefont
  {Mart\'{i}nez-Rat\'{o}n}}\ and\ \bibinfo {author} {\bibfnamefont {J.~A.}\
  \bibnamefont {Cuesta}},\ }\href@noop {} {\bibfield  {journal} {\bibinfo
  {journal} {J. Chem. Phys}\ }\textbf {\bibinfo {volume} {118}},\ \bibinfo
  {pages} {10164} (\bibinfo {year} {2003})}\BibitemShut {NoStop}%
\bibitem [{\citenamefont {van Roij}\ \emph {et~al.}(2000)\citenamefont {van
  Roij}, \citenamefont {Dijkstra},\ and\ \citenamefont {Evans}}]{roij2000}%
  \BibitemOpen
  \bibfield  {author} {\bibinfo {author} {\bibfnamefont {R.}~\bibnamefont {van
  Roij}}, \bibinfo {author} {\bibfnamefont {M.}~\bibnamefont {Dijkstra}}, \
  and\ \bibinfo {author} {\bibfnamefont {R.}~\bibnamefont {Evans}},\
  }\href@noop {} {\bibfield  {journal} {\bibinfo  {journal} {Euro. Phys.
  Lett.}\ }\textbf {\bibinfo {volume} {49}},\ \bibinfo {pages} {350} (\bibinfo
  {year} {2000})}\BibitemShut {NoStop}%
\bibitem [{\citenamefont {Ramirez-Pastor}\ \emph {et~al.}(1999)\citenamefont
  {Ramirez-Pastor}, \citenamefont {Eggarter}, \citenamefont {Pereyra},\ and\
  \citenamefont {Riccardo}}]{pastor1999}%
  \BibitemOpen
  \bibfield  {author} {\bibinfo {author} {\bibfnamefont {A.~J.}\ \bibnamefont
  {Ramirez-Pastor}}, \bibinfo {author} {\bibfnamefont {T.~P.}\ \bibnamefont
  {Eggarter}}, \bibinfo {author} {\bibfnamefont {V.~D.}\ \bibnamefont
  {Pereyra}}, \ and\ \bibinfo {author} {\bibfnamefont {J.~L.}\ \bibnamefont
  {Riccardo}},\ }\href@noop {} {\bibfield  {journal} {\bibinfo  {journal}
  {Phys. Rev. B}\ }\textbf {\bibinfo {volume} {59}},\ \bibinfo {pages} {11027}
  (\bibinfo {year} {1999})}\BibitemShut {NoStop}%
\bibitem [{\citenamefont {Rom\'{a}}\ \emph {et~al.}(2003)\citenamefont
  {Rom\'{a}}, \citenamefont {Ramirez-Pastor},\ and\ \citenamefont
  {Riccardo}}]{pastor2003}%
  \BibitemOpen
  \bibfield  {author} {\bibinfo {author} {\bibfnamefont {F.}~\bibnamefont
  {Rom\'{a}}}, \bibinfo {author} {\bibfnamefont {A.~J.}\ \bibnamefont
  {Ramirez-Pastor}}, \ and\ \bibinfo {author} {\bibfnamefont {J.~L.}\
  \bibnamefont {Riccardo}},\ }\href@noop {} {\bibfield  {journal} {\bibinfo
  {journal} {Langmuir}\ }\textbf {\bibinfo {volume} {19}},\ \bibinfo {pages}
  {6770} (\bibinfo {year} {2003})}\BibitemShut {NoStop}%
\bibitem [{\citenamefont {Bellemans}\ and\ \citenamefont
  {Nigam}(1967)}]{nigam1967}%
  \BibitemOpen
  \bibfield  {author} {\bibinfo {author} {\bibfnamefont {A.}~\bibnamefont
  {Bellemans}}\ and\ \bibinfo {author} {\bibfnamefont {R.~K.}\ \bibnamefont
  {Nigam}},\ }\href@noop {} {\bibfield  {journal} {\bibinfo  {journal} {J.
  Chem. Phys.}\ }\textbf {\bibinfo {volume} {46}},\ \bibinfo {pages} {2922}
  (\bibinfo {year} {1967})}\BibitemShut {NoStop}%
\bibitem [{\citenamefont {Pearce}\ and\ \citenamefont
  {Seaton}(1988)}]{pearce1988}%
  \BibitemOpen
  \bibfield  {author} {\bibinfo {author} {\bibfnamefont {P.~A.}\ \bibnamefont
  {Pearce}}\ and\ \bibinfo {author} {\bibfnamefont {K.~A.}\ \bibnamefont
  {Seaton}},\ }\href@noop {} {\bibfield  {journal} {\bibinfo  {journal} {J.
  Stat. Phys.}\ }\textbf {\bibinfo {volume} {53}},\ \bibinfo {pages} {1061}
  (\bibinfo {year} {1988})}\BibitemShut {NoStop}%
\bibitem [{\citenamefont {Ramola}\ and\ \citenamefont
  {Dhar}(2012)}]{kabir2012}%
  \BibitemOpen
  \bibfield  {author} {\bibinfo {author} {\bibfnamefont {K.}~\bibnamefont
  {Ramola}}\ and\ \bibinfo {author} {\bibfnamefont {D.}~\bibnamefont {Dhar}},\
  }\href@noop {} {\bibfield  {journal} {\bibinfo  {journal} {Phys. Rev. E}\
  }\textbf {\bibinfo {volume} {86}},\ \bibinfo {pages} {031135} (\bibinfo
  {year} {2012})}\BibitemShut {NoStop}%
\bibitem [{\citenamefont {Baxter}(1980)}]{baxter1980}%
  \BibitemOpen
  \bibfield  {author} {\bibinfo {author} {\bibfnamefont {R.~J.}\ \bibnamefont
  {Baxter}},\ }\href@noop {} {\bibfield  {journal} {\bibinfo  {journal} {J.
  Phys. A}\ }\textbf {\bibinfo {volume} {13}},\ \bibinfo {pages} {L61}
  (\bibinfo {year} {1980})}\BibitemShut {NoStop}%
\bibitem [{\citenamefont {Heilmann}\ and\ \citenamefont
  {Praestgaard}(1973)}]{heilmann1973}%
  \BibitemOpen
  \bibfield  {author} {\bibinfo {author} {\bibfnamefont {O.~J.}\ \bibnamefont
  {Heilmann}}\ and\ \bibinfo {author} {\bibfnamefont {E.}~\bibnamefont
  {Praestgaard}},\ }\href@noop {} {\bibfield  {journal} {\bibinfo  {journal}
  {J. Stat. Phys}\ }\textbf {\bibinfo {volume} {9}},\ \bibinfo {pages} {23}
  (\bibinfo {year} {1973})}\BibitemShut {NoStop}%
\bibitem [{\citenamefont {Dickman}(2012)}]{dickman2012}%
  \BibitemOpen
  \bibfield  {author} {\bibinfo {author} {\bibfnamefont {R.}~\bibnamefont
  {Dickman}},\ }\href@noop {} {\bibfield  {journal} {\bibinfo  {journal} {J.
  Chem. Phys.}\ }\textbf {\bibinfo {volume} {136}},\ \bibinfo {pages} {174105}
  (\bibinfo {year} {2012})}\BibitemShut {NoStop}%
\bibitem [{\citenamefont {Verberkmoes}\ and\ \citenamefont
  {Nienhuis}(1999)}]{nienhuis1999}%
  \BibitemOpen
  \bibfield  {author} {\bibinfo {author} {\bibfnamefont {A.}~\bibnamefont
  {Verberkmoes}}\ and\ \bibinfo {author} {\bibfnamefont {B.}~\bibnamefont
  {Nienhuis}},\ }\href@noop {} {\bibfield  {journal} {\bibinfo  {journal}
  {Phys. Rev. Lett.}\ }\textbf {\bibinfo {volume} {83}},\ \bibinfo {pages}
  {3986} (\bibinfo {year} {1999})}\BibitemShut {NoStop}%
\bibitem [{\citenamefont {Kundu}\ \emph {et~al.}(2012)\citenamefont {Kundu},
  \citenamefont {Rajesh}, \citenamefont {Dhar},\ and\ \citenamefont
  {Stilck}}]{joyjit_dae}%
  \BibitemOpen
  \bibfield  {author} {\bibinfo {author} {\bibfnamefont {J.}~\bibnamefont
  {Kundu}}, \bibinfo {author} {\bibfnamefont {R.}~\bibnamefont {Rajesh}},
  \bibinfo {author} {\bibfnamefont {D.}~\bibnamefont {Dhar}}, \ and\ \bibinfo
  {author} {\bibfnamefont {J.~F.}\ \bibnamefont {Stilck}},\ }\href@noop {}
  {\bibfield  {journal} {\bibinfo  {journal} {AIP Conf. Proc.}\ }\textbf
  {\bibinfo {volume} {1447}},\ \bibinfo {pages} {113} (\bibinfo {year}
  {2012})}\BibitemShut {NoStop}%
\bibitem [{\citenamefont {Matoz-Fernandez}\ \emph
  {et~al.}(2008{\natexlab{b}})\citenamefont {Matoz-Fernandez}, \citenamefont
  {Linares},\ and\ \citenamefont {Ramirez-Pastor}}]{fernandez2008c}%
  \BibitemOpen
  \bibfield  {author} {\bibinfo {author} {\bibfnamefont {D.~A.}\ \bibnamefont
  {Matoz-Fernandez}}, \bibinfo {author} {\bibfnamefont {D.~H.}\ \bibnamefont
  {Linares}}, \ and\ \bibinfo {author} {\bibfnamefont {A.~J.}\ \bibnamefont
  {Ramirez-Pastor}},\ }\href@noop {} {\bibfield  {journal} {\bibinfo  {journal}
  {J. Chem. Phys.}\ }\textbf {\bibinfo {volume} {128}},\ \bibinfo {pages}
  {214902} (\bibinfo {year} {2008}{\natexlab{b}})}\BibitemShut {NoStop}%
\bibitem [{\citenamefont {DiMarzio}(1961)}]{dimarzio1961}%
  \BibitemOpen
  \bibfield  {author} {\bibinfo {author} {\bibfnamefont {E.}~\bibnamefont
  {DiMarzio}},\ }\href@noop {} {\bibfield  {journal} {\bibinfo  {journal} {J.
  Chem. Phys.}\ }\textbf {\bibinfo {volume} {35}},\ \bibinfo {pages} {658}
  (\bibinfo {year} {1961})}\BibitemShut {NoStop}%
\bibitem [{\citenamefont {Sokolova}\ and\ \citenamefont
  {Tumanyan}(2000)}]{sokolova2000}%
  \BibitemOpen
  \bibfield  {author} {\bibinfo {author} {\bibfnamefont {E.~P.}\ \bibnamefont
  {Sokolova}}\ and\ \bibinfo {author} {\bibfnamefont {N.~P.}\ \bibnamefont
  {Tumanyan}},\ }\href@noop {} {\bibfield  {journal} {\bibinfo  {journal} {Liq.
  Crys.}\ }\textbf {\bibinfo {volume} {27}},\ \bibinfo {pages} {813} (\bibinfo
  {year} {2000})}\BibitemShut {NoStop}%
\end{thebibliography}

%

\end{document}